\documentclass[3p]{elsarticle}

\journal{Applied Mathematical Modelling}

\usepackage[colorlinks,linkcolor=red,anchorcolor=blue,citecolor=green,CJKbookmarks=True]{hyperref}
\usepackage{graphicx}      
\usepackage{natbib}        

\usepackage{amsmath,amssymb,amsfonts}
\usepackage{algorithm}
\usepackage{algorithmic}

\usepackage{savesym}
\savesymbol{AND}

\usepackage{textcomp}
\usepackage{url}
\usepackage{color}
\usepackage{tikz}
\usetikzlibrary{arrows,shapes,chains}  	
\usepackage{xcolor}
\usepackage{lipsum}
\usepackage{epstopdf}
\usepackage{float}
\usepackage{amsopn}

\usepackage{enumerate}
\usepackage{cancel}
\usepackage{mathrsfs}
\usepackage{mathdots}
\usepackage{euscript}
\usepackage{amscd}
\usepackage{placeins}

\usepackage{multirow} 
\usepackage{xcolor}
\usepackage{booktabs}
\usepackage{mathtools}

\newtheorem{problem}{Problem}
\newtheorem{remark}{Remark}

\newdefinition{rmk}{Remark}
\newproof{pf}{Proof}

\newcommand{\bmat}{\left[ \begin{matrix}}
	\newcommand{\emat}{\end{matrix} \right]}
\newcommand{\innerprod}[2]{\langle{#1},\,{#2}\rangle}
\DeclareMathOperator{\E}{{\mathbb E}}
\newcommand{\Rbb}{\mathbb R}

\newcommand{\Cbb}{\mathbb C}

\newcommand{\Zbb}{\mathbb Z}

\newcommand{\Tbb}{\mathbb T}

\newcommand{\xb}{\mathbf  x}
\newcommand{\yb}{\mathbf  y}
\newcommand{\sbf}{\mathbf  s}  
\newcommand{\zb}{\mathbf  z}
\newcommand{\wb}{\mathbf  w}

\newcommand{\eb}{\mathbf  e}

\newcommand{\ab}{\mathbf a}
\newcommand{\bb}{\mathbf  b}

\newcommand{\qb}{\mathbf q}
\newcommand{\tb}{\mathbf t}
\newcommand{\kb}{\mathbf k}

\newcommand{\zerob}{\mathbf 0}

\newcommand{\Cb}{\mathbf C}

\newcommand{\Mb}{\mathbf M}
\newcommand{\Nb}{\mathbf N}

\newcommand{\Tb}{\mathbf T}

\newcommand{\thetab}{\boldsymbol{\theta}}

\newcommand{\sigmab}{\boldsymbol{\sigma}}

\newcommand{\Sigmab}{\boldsymbol{\Sigma}}

\newcommand{\taub}{\boldsymbol{\tau}}
\newcommand{\mub}{\boldsymbol{\mu}}

\newcommand{\Ncal}{\mathcal{N}}
\newcommand{\Fcal}{\mathcal{F}}

\renewcommand{\d}{\mathrm{d}}

\newcommand{\srm}{\mathrm{s}}

\renewcommand{\Re}{\mathrm{Re}}

\begin{document}

\begin{frontmatter}

\title{Sampling Gaussian Stationary Random Fields:\\ A Stochastic Realization Approach}
\tnotetext[mytitlenote]{Partial results of this paper will be presented in the 23rd Chinese Conference on System Simulation Technology and its Application (CCSSTA 2022).}


\author[a]{Bin Zhu\corref{cor1}}
\ead{zhub26@mail.sysu.edu.cn}
\cortext[cor1]{Corresponding author}

\author[a]{Jiahao Liu}
\ead{liujh226@mail2.sysu.edu.cn}

\author[a]{Zhengshou Lai}
\ead{laizhengsh@mail.sysu.edu.cn}

\author[b]{Tao Qian}
\ead{tqian@must.edu.mo}

\address[a]{School of Intelligent Systems Engineering, Sun Yat-sen University, Gongchang Road 66, 518107 Shenzhen, China}
\address[b]{Macau Centre for Mathematical Sciences, Macau University of Science and Technology, Macau, China}

\begin{abstract}
	Generating large-scale samples of stationary random fields is of great importance in the fields such as geomaterial modeling and uncertainty quantification.
	Traditional methodologies based on covariance matrix decomposition have the difficulty of being computationally expensive, which is even more serious when the dimension of the random field is large.
	This paper proposes an efficient stochastic realization approach for sampling Gaussian stationary random fields from a systems and control point of view. 
	Specifically, we take the exponential and Gaussian covariance functions as examples and make a decoupling assumption when there are multiple dimensions. 
	Then a rational spectral density is constructed in each dimension using techniques from covariance extension, and the corresponding autoregressive moving-average (ARMA) model is obtained via spectral factorization.
	As a result, samples of the random field with a specific covariance function can be generated very efficiently in the space domain by implementing the ARMA recursion using a white noise input. 
	Such a procedure is computationally cheap due to the fact that the constructed ARMA model has a low order. 
	Furthermore, the same method is integrated to multiscale simulations where interpolations of the generated samples are achieved when one zooms into finer scales.
	Both theoretical analysis and simulation results show that our approach performs favorably compared with covariance matrix decomposition methods. 
\end{abstract}


\end{frontmatter}

\section{Introduction}\label{Sec:Introduction}


There is a vast number of applications of Gaussian random fields across several engineering and scientific disciplines including systems and control \cite{ringh2016multidimensional,zhu2021multidimensional}, signal processing \cite{stoica2005spectral,qian2022afd}, geotechnical engineering \citep{chen2016cpt,latz2019fast}, image processing \cite{winkler2003image}, biology, and meteorology \cite{croci2018efficient,khristenko2019analysis}. 
In these applications, we often face the standard problem of sampling a random field which could be used e.g., for the numerical solution of a stochastic partial differential equation (PDE). 
In particular in geotechnical engineering, certain geomaterial properties are modeled as zero-mean stationary random fields indexed by spatial or temporal variables. Then the correlations between random variables in the field are described by the covariance function. 
In order to carry out simulations in geomaterial modeling, a first step is to generate (possibly large-scale) samples of a stationary random field such that its covariances coincide with the values of a prescribed covariance function. 
A traditional method for this problem is called covariance matrix decomposition (CMD) \cite{fenton2008risk} which employs e.g., the Cholesky factorization.
Such a method in general costs $O(N^3)$ flops given an $N\times N$ matrix, which is computationally prohibitive when the covariance matrix has a large size. The latter case is typical for multidimensional random fields. For example, a $3$-d random field with a (moderate) size $100\times 100\times 100$ results in a $10^6\times 10^6$ covariance matrix after vectorization. Thus applications of CMD are seriously limited to small-scale and unidimensional cases (time series). However, many practical problems involve multidimensional random fields \cite{vanmarcke2010random,graham2018analysis} and the ability to efficiently generate large-scale samples is also important.  

In the literature, there are a number of methods to handle such a problem. Recent developments include \citet{graham2018analysis} which aims to embed a finite multilevel Toeplitz covariance matrix into a larger positive definite multilevel circulant matrix, following earlier works on the embedding problem for (one-level) Toeplitz covariance matrices \cite{dembo1989embedding,dietrich1997fast}. Then the fast Fourier transform (FFT) can be used to compute the spectral decomposition of the (multilevel) circulant matrix at a reduced cost.  
Another direction is to use the Karhunen--Lo{\`e}ve expansion for the continuous covariance kernel and to truncate the expansion for practical computations \cite{zheng2017simulation,latz2019fast,qian2022afd}.
A different strategy for the approximation of the covariance matrix involves the notion of $H$ matrices \cite{feischl2018fast} of which the square root can be computed at an almost linear cost.



Although the above works are mathematically interesting and deal with general covariance functions, difficulties can still arise when one wants to generate very large-scale samples, especially when the dimension is greater than or equal to three. In order to handle the latter issue, the paper \cite{li2019stepwise} made a \emph{decoupling assumption} on the multivariable covariance function and proposed a stepwise CMD method which essentially computes the matrix square root along each dimension and thus reduces the computational cost compared to a full CMD. However, the space domain computation in that paper can be once again greatly reduced if one is able to recognize the frequency-domain structure of the covariance function, namely the \emph{spectral density} of the random field. This is the main idea behind the current paper. More specifically, we draw inspiration from the systems and control literature on \emph{stochastic realization} \cite{LP15} and \emph{rational covariance extension} in which one aims to describe an underlying random process with a linear dynamical system driven by white noise. The latter topic has undergone decades of development from scalar random processes to vector random fields, see \cite{BGL-98,byrnes2001finite,enqvist2004aconvex,Georgiou-06,GL-08,FPR-08,LPcirculant-13,Zhu-Baggio-19,zhu2020well,ZFKZ2-M2-Aut,Liu-Zhu-2021,Liu-Zhu-2022-CCSSTA} and the references therein. 
In particular, we have shown that the exponential covariance function corresponds exactly to an autoregressive (AR) model (a rational filter) of order one, which is easy to implement recursively and permits the sample generation at a \emph{linear} cost. The decoupling assumption makes straightforward the generalization to multidimensional random fields, and the resulting multidimensional filter is simply a product of individual filters in each dimension. In this way, the filtering procedure is also decoupled as expected.



In addition, our approach is extremely suitable for multiscale simulations, see e.g., \citet {chen2016cpt}, where the generated samples of a random field are \emph{interpolated} and fed into a numerical PDE solver in order to obtain a refined solution.
The usual interpolation method samples the probability density of the fine-scale random variables whose values are to be determined, conditioned on the coarse-scale samples that have already been generated. The advantage of our approach is that, once a suitable refined noise input has been determined, only ``boundary'' samples that are necessary to initiate the \emph{fine-scale} ARMA recursions need to be computed from the conditional probability density, and the rest fine-scale samples are generated in the same fashion as the coarse-scale sampling. 
 

The outline of the paper is as follows. In Section~\ref{Sec:Problem statement} we state the problem of sampling a stationary random field with a given covariance function that can be decoupled in each dimension. 
In Section~\ref{Sec:Spectral analysis} we propose a stochastic realization approach to the sampling problem and focus on the solutions for the exponential and Gaussian covariance functions.
In Section~\ref{Sec:Multi-scale random fields} we integrate our method to multiscale simulations and provide an explicit solution procedure for the bivariate exponential covariance function.
A number of numerical simulations are presented in Section~\ref{Sec:Numerical examples}, and in Section~\ref{Sec:Conclusions} we draw the conclusions.

\section{Background on sampling random fields}\label{Sec:Problem statement}

Let $y(\tb,\omega)$ be a $d$-dimensional real random field over a probability space $(\Omega, \Fcal, P)$ where $\tb=(t_1,\dots,t_d)\in\Rbb^d$ can be interpreted as a space (or spatio-temporal) coordinate vector. For each fixed $\tb\in\Rbb^d$, assume that $y(\tb,\cdot)$ is a zero-mean real-valued random variable with a finite variance, that is,
\begin{equation*}
\E y(\tb,\cdot) = 0\quad \text{and} \quad \E\, [y(\tb,\cdot)]^2 < \infty
\end{equation*}
where $\E$ indicates mathematical expectation.
It is customary to suppress the dependence on $\omega$ and write simply $y(\tb)$. Assume further that the random field under consideration is \emph{stationary}, which means that the covariance function
\begin{equation*}
\rho(\tb,\sbf) := \E [y(\tb)y(\sbf)]
\end{equation*}
depends only on the difference $\taub:=\tb-\sbf$ between the arguments, so we can write $\rho(\taub)$ instead. Notice the symmetry $\rho(-\taub) = \rho(\taub)$.

In applications of geotechnical engineering, see e.g., \cite{firouzianbandpey2015effect,ching2017characterizing}, it is often assumed that the covariance function has a decoupled form, 
namely
\begin{equation}\label{eq:decoupled form}
	\rho(\taub)=\rho_1(\tau_1) \rho_2(\tau_2) \cdots \rho_d(\tau_d),
\end{equation}
where each $\rho_j$ is a covariance function in one variable, and $\tau_j\in \Rbb$ is the $j$-th component of $\taub$. 
In the following, since we shall be concerned with the sampling problem of the random field $y(\tb)$, let us now define the sampled version of the random field as well as the covariance function.
Take $\tau_j = x_j T_j$ where $T_j>0$ is sampling distance and $x_j\in\Zbb$. Define a random field on the integer grid $\Zbb^d$ via
\begin{equation}\label{sampled_random_field}
y_{\srm}(\xb) = y(x_1 T_1,\dots,x_d T_d)
\end{equation}
where the subscript s means ``sampled''. Then it is easy to deduce that the covariance function of the discrete random field $y_{\srm}$ is
\begin{equation}\label{rho_sampled}
	\rho_{\srm}(\kb) = \rho(k_1 T_1,\dots,k_d T_d) = \rho_1 (k_1 T_1) \cdots \rho_d (k_d T_d),
\end{equation}
where $\kb=(k_1,\dots,k_d)\in\Zbb^d$ denotes the difference between two discrete grid points.
The sampling problem can then be phrased as follows.
\begin{problem}\label{prob_sampling_general}
	Given a covariance function $\rho(\taub)$ of the form \eqref{eq:decoupled form}, a vector $\Tb=(T_1,\dots,T_d)$ of sampling distances, and a vector $\Nb=(N_1,\dots,N_d)$ of positive integers, generate samples of the random field $y_{\srm}(\xb)$ in \eqref{sampled_random_field} for $\xb$ in the index set $($a regular cuboid$)$
	\begin{equation}\label{ind_set_RF}
	\Zbb^d_{\Nb} := \{(x_1,\dots,x_d) : 0\leq x_j\leq N_j-1,\ j=1,\dots,d\}
	\end{equation}
	such that its covariance function coincides with the sampled version $\rho_{\srm}(\kb)$ in \eqref{rho_sampled}.
\end{problem}

The most straightforward approach for this problem is covariance matrix decomposition mentioned in the Introduction. More precisely, since the index set \eqref{ind_set_RF} has a finite cardinality, all the samples can be stacked into a long vector $\yb$ of dimension $|\Nb|:=\prod_{j=1}^{d}N_j$. Then the covariance matrix $\Sigmab:=\E(\yb\yb^\top)$ could in principle be evaluated elementwise according to \eqref{rho_sampled}. Problem~\ref{prob_sampling_general} would then be solved via simple linear algebra
\begin{equation}\label{eq:Ranom field realization}
	\yb=L\wb,
\end{equation}
where $\wb\sim \Ncal(\mathbf{0},I)$ is an i.i.d.~standard normal random vector of dimension $|\Nb|$, and $L$ is any matrix square root of $\Sigmab$ which can in particular, be taken as the Cholesky factor such that $LL^{\top}=\Sigmab$. However, it is well known that the matrix factorization, which is the major computational burden here, involves $O(|\Nb|^3)$ flops. Therefore, such a naive approach works only for random fields with a dimension $d=1$ or $2$ when the size of the samples $|\Nb|$ is not too large. For the generation of large-scale samples, one needs to exploit the inherent structure of the covariance function in order to facilitate fast computation.
The latter point is indeed the theme of the next section where we will propose an efficient stochastic realization 
approach to Problem~\ref{prob_sampling_general}.
More specifically, we consider two types of $1$-d covariance functions of practical interest:
\begin{enumerate}
	\item the exponential type 
	\begin{equation}\label{eq:exponential function}
		\rho(x) = \sigma^2 e^{-\alpha|x|},
	\end{equation}

	\item the Gaussian type 
	\begin{equation}\label{eq:Gaussian function}
		\rho(x) = \sigma^2 e^{-\alpha|x|^{2}},
	\end{equation}
\end{enumerate}
where $\sigma^2$ is the variance of the random field
and $\alpha>0$ is a parameter. The multidimensional covariance function is formed through the product in \eqref{eq:decoupled form}.
Notice that our method works also for other types of covariance functions $\rho(x)$ provided that the decoupling assumption \eqref{eq:decoupled form} holds.
In this case, the spectral density of the random field can be well approximated in each dimension by a low-order rational model.


\begin{remark}
The exponential and the Gaussian covariance functions are special cases of the Mat\'{e}rn family of covariance functions, defined as
	\begin{equation}\label{matern_cov}
	\rho(\taub) = \kappa(\|\taub\|/\lambda)\quad \text{where}\quad \kappa(r) = \sigma^2 \frac{2^{1-\nu}}{\Gamma(\nu)} (\sqrt{2\nu}\, r)^\nu K_{\nu}(\sqrt{2\nu}\, r).
	\end{equation}
	In the formulas above, $\taub\in\Rbb^d$, $\|\cdot\|$ is the Euclidean norm, $\lambda$ the correlation length, $\sigma^2$ the variance, $\nu>0$ a smoothness parameter, $\Gamma$ the gamma function, and $K_{\nu}$ the modified Bessel function of the second kind. Notice that the case $\nu=1/2$ corresponds to the exponential covariances, while $\nu=\infty$ corresponds to the Gaussian function of the form $\kappa(r) = \sigma^2 e^{-r^2/2}$. See e.g., \cite[Example 2.7]{graham2018analysis}.
\end{remark}

\section{A stochastic realization approach}\label{Sec:Spectral analysis}

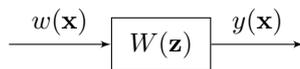
\begin{figure}[h]
	\centering
	\tikzstyle{int}=[draw, minimum size=2em]
	\tikzstyle{init} = [pin edge={to-,thin,black}]
	\begin{tikzpicture}[node distance=2cm,auto,>=latex']
	\node [int] (a) {$\ W(\zb)\ $};
	\node (b) [left of=a, coordinate] {};
	\node (c) [right of=a] {};
	\path[->] (b) edge node {$w(\xb)$} (a);
	\path[->] (a) edge node {$y(\xb)$} (c);
	\end{tikzpicture}
	\caption{A $d$-dimensional linear stochastic system with a white noise input.}
	\label{Fig:d_linear_system}
\end{figure}

Let $\zb=(z_1,\cdots,z_d)$ be a vector of indeterminates.
Consider a $d$-dimensional discrete-``time'' linear stochastic system as depicted in Fig.~\ref{Fig:d_linear_system}, where 
\begin{equation}
W(\zb) = \sum_{\kb \in \Zbb^d} \gamma(\kb) \,\zb^{-\kb}
\end{equation}
is the transfer function, also called a shaping filter in the signal processing literature. Here the function $\gamma:\Zbb^d\to \Rbb$ is called the \emph{impulse response} of the system and $\zb^{\kb}$ is a shorthand notation for $z_1^{k_1}\cdots z_d^{k_d}$. Moreover, the symbol $\zb^{-\kb}$ can be interpreted as a $\kb$-step delay operator. The system is excited by a normalized white noise $w(\xb)$ such that for any $\xb\in\Zbb^d$,
\begin{equation}
\E [w(\xb)]=0\quad\text{and} \quad \E[w(\xb+\kb) w(\xb)]=\delta_{\kb,\zerob} 
= \begin{cases}
1 & \text{if}\ \kb=\zerob, \\
0 & \text{otherwise}.
\end{cases}
\end{equation}
The output $y(\xb)$ is a zero-mean stationary random field. Symbolically, we write
\begin{equation}\label{eq:linear-stochastic-realization}
y(\xb)=W(\zb)w(\xb):= \sum_{\kb \in \Zbb^d} \gamma_{\kb} w(\xb-\kb).
\end{equation}
Notice that this is a standard model for stationary processes which goes back to the prediction theory of Wiener in the 1940s.


Let $\sigma(\kb):=\E [y(\xb+\kb) y(\xb)]$ be the covariance function of $y(\xb)$ and let $\Tbb:=[-\pi, \pi)$ denote the frequency interval. Then the spectral density of $y(\xb)$ is by definition \cite{stoica2005spectral,LP15} the multidimensional discrete-time Fourier transform (DTFT) of the covariance function:
\begin{equation}
\Phi(e^{i\thetab})=\sum_{\kb \in \Zbb^d} \sigma(\kb)\, e^{-i \innerprod{\kb}{\thetab}},
\end{equation} 
where the frequency vector $\thetab=(\theta_1,\dots,\theta_d)\in \Tbb^d$, $e^{i\thetab}:=(e^{i\theta_1},\dots,e^{i\theta_d})$ is a point on the $d$-torus (which is isomorphic to $\Tbb^d$), and
$\innerprod{\kb}{\thetab}:=k_1\theta_1+\cdots+k_d\theta_d$ is the standard inner product in $\Rbb^d$. It then follows from the spectral theory of stationary random fields \cite{yaglom1957some} that
\begin{equation}\label{eq:Phi=w(z)w(z^-1)}
\Phi(e^{i\thetab})=|W(e^{i\thetab})|^2
\end{equation}
where $W(e^{i\thetab})= \sum_{\kb \in \Zbb^d} \gamma(\kb) \,e^{-i\innerprod{\kb}{\thetab}}$, so $\Phi(e^{i\thetab})$ takes \emph{nonnegative} values.
On the other hand, if the spectral density $\Phi$ satisfies certain analytic properties, then it admits a \emph{spectral factor} $W$, see \cite{LP15}.



We are mostly interested in the case where $W(\zb)$ is a \emph{rational} function, that is, it can be expressed as a ratio of two polynomials:
\begin{equation}\label{eq:transfer_func}
	W(\zb)=\frac{b(\zb)}{a(\zb)}=\frac{\sum_{\kb \in \Lambda_{+,2}}b_{\kb}\zb^{-\kb}}{\sum_{\kb \in \Lambda_{+,1}}a_{\kb}\zb^{-\kb}},
\end{equation}
where
\begin{equation*}
\begin{aligned}
\Lambda_{+,1} & := \{(k_1,\cdots,k_d)\in\Zbb^d : 0\leq k_j \leq m_j,\ j=1, \cdots, d \}, \\
\Lambda_{+,2} & := \{(k_1,\cdots,k_d)\in\Zbb^d : 0\leq k_j \leq n_j,\ j=1,\cdots,d\}
\end{aligned}
\end{equation*}
are two index sets with positive integers $m_j, n_j$ given for $j=1,\dots,d$.
Then the system \eqref{eq:linear-stochastic-realization} can equivalently be described in the time domain as an autoregressive moving-average (ARMA) model
\begin{equation}\label{eq:ARMA recursion}
	\sum_{\kb \in \Lambda_{+,1}} a_{\kb}\, y(\xb-\kb) = \sum_{\kb \in \Lambda_{+,2}} b_{\kb}\, w(\xb-\kb).
\end{equation}
Such a model is extremely useful in practice because rational functions (in fact, polynomials) can approximate any continuous function if the model order is sufficiently large. If the moving-average part of the model is trivial, i.e., $b(\zb)\equiv b_{\zerob}$ is a constant, then \eqref{eq:ARMA recursion} reduces to a simpler AR model. 


In the above context, Problem~\ref{prob_sampling_general} can be posed more concretely as follows:

\begin{problem}\label{prob_realization}	
	Given a sampled covariance function $\rho_{\srm}(\kb)$ in \eqref{rho_sampled}, find a rational filter $W(\zb)$ of the form \eqref{eq:transfer_func} such that when it is fed with a normalized white noise, the covariance function of the output $y(\xb)$ coincides with $\rho_{\srm}(\kb)$.
	Equivalently, we seek a rational spectral density satisfying the trigonometric moment equations
	\begin{equation}\label{d-dim_moment_eqns}
	\int_{\Tbb^d} e^{i\innerprod{\kb}{\thetab}} \Phi(e^{i\thetab}) \d\mu(\thetab) = \rho_{\srm}(\kb)\quad \forall \kb\in\Zbb^d
	\end{equation}
	where $\d\mu(\thetab) = \frac{1}{(2\pi)^d} \prod_{j=1}^{d} \d\theta_j$ is the normalized Lebesgue on $\Tbb^d$.
\end{problem}

Notice that the equivalence above is understood modulo the spectral factorization \eqref{eq:Phi=w(z)w(z^-1)}.
Once the above problem is solved, samples of the random field $y_{\srm}(\xb)$ for $\xb$ indexed in \eqref{ind_set_RF} can be generated efficiently via the ARMA recursion \eqref{eq:ARMA recursion} with a white noise input.



Next, in view of the decoupling assumption \eqref{rho_sampled}, it follows easily from the multidimensional DTFT that the corresponding spectral density ${\Phi}(e^{i\thetab})$ also has a decoupled form 
\begin{equation}\label{Spe-deco}
	{\Phi} (e^{i\thetab}) = {\Phi}_1(e^{i\theta_1})\, \cdots\, {\Phi}_d(e^{i\theta_d}),
\end{equation}
where the factor ${\Phi}_j(e^{i\theta_j})$ can be interpreted as the spectral density in the $j$-th dimension for $j=1,\cdots,d$, i.e., it is the $1$-d DTFT of the sampled covariance function $\rho_{\srm,j}(k_j):=\rho_j (k_j T_j)$.
Therefore, the $d$-dimensional moment equations \eqref{d-dim_moment_eqns} decouple into $d$ sets of unidimensional moment equations
\begin{equation}\label{one-d_moment_eqns}
\int_{\Tbb} e^{ik_j\theta_j} \Phi_j(e^{i\theta_j}) \frac{\d\theta_j}{2\pi} = \rho_{\srm,j}(k_j) \quad \forall k_j\in\Zbb,\ j=1,\dots,d,
\end{equation}
whose solutions have been extensively studied in the literature.
After each $\Phi_j$ has been constructed from the covariance function $\rho_{\srm,j}$, we can perform the spectral factorization $\Phi_j (z_j) = W_j (z_j) W_j (z_j^{-1})$ to obtain the transfer function $W_j(z_j)$. Notice that since $\Phi_j$ is constrained to be rational, the spectral factorization reduces to that for positive trigonometric polynomials, for which there are a number of algorithms \cite{sayed2001survey}.
Hence, the above procedure leads to a $d$-dimensional ARMA model
\begin{equation}\label{d-dimensional ARMA model}
	y(\xb) = \underbrace{W_1(z_1) \, \cdots\, W_d(z_d)}_{=: W(\zb)} \, w(\xb)
\end{equation}
again in a decoupled form. 
The main steps of our approach for sampling stationary random fields are summarized as follows.
\begin{enumerate}
	\item[1)] Given a sampled covariance function \eqref{rho_sampled}, solve the decoupled moment equations \eqref{one-d_moment_eqns} for a rational spectral density of the form \eqref{Spe-deco}.
	\item[2)] Do spectral factorization to obtain the linear filter in \eqref{d-dimensional ARMA model}.
	\item[3)] Feed the filter with a Gaussian i.i.d.~white noise and collect the output random field.
\end{enumerate}


In the next subsections, we focus on the nontrivial Step 1 with two types of covariance functions mentioned before, i.e., the exponential and Gaussian covariance functions. It is worth remarking that the case of the exponential covariance function admits an \emph{exact} rational spectrum and hence a shaping filter in a \emph{closed form}.
Although the Gaussian covariance function does not lead to an analytic solution, it can be well \emph{approximated} in the frequency domain by a rational spectral density in the sense that only low-order covariances with significant values are matched in \eqref{one-d_moment_eqns}. 


\subsection{Solution for the exponential covariance function: an AR$(1)$ model}\label{subsec-expon}


Under the decoupling assumption for the $d$-dimensional covariance function, we only need to solve Problem~\ref{prob_realization} for each sampled covariance function $\rho_{\srm,j}(k_j)=\rho_j (k_j T_j)$ of one variable, as discussed previously. Hence, we suppress the subscript $j$.
Consider first the exponential covariance function in \eqref{eq:exponential function}, i.e., $\rho(x) = \sigma^2 e^{-\alpha|x|}$ with $x\in\Rbb$. We can for simplicity take $\sigma^2=1$ since it is only a multiplicative constant. Let $T>0$ be the sampling distance, and we have
\begin{equation}\label{sampled_cov_exp}
\rho_{\srm}(k) := \rho(k T) = e^{-\alpha T|k|},\quad k\in \Zbb.
\end{equation}
Let $r=e^{-\alpha T}$.
Using the DTFT pair
\begin{equation}
	x(k) = r^{k}u(k) \xmapsto{\Fcal} X(e^{i\theta}) = \frac{1}{1 - r e^{-i\theta}} 
\end{equation}
with $0<|r|<1$ and $u(k)$ the discrete-time unit step function, the spectral density corresponding to $\rho_\srm(k)$ is
\begin{equation}\label{Phi_posi_real}
	{\Phi}(e^{i\theta}) = Z(e^{i\theta}) + Z(e^{i\theta})^* = 2\Re\{ Z(e^{i\theta})\},
\end{equation}
where $^*$ denotes complex conjugate and
\begin{equation}
	Z(e^{i\theta}) = \frac{1}{1 - e^{-\alpha T} e^{-i\theta}} - \frac{1}{2}
\end{equation}
is the so-called \emph{positive real} part of ${\Phi} (e^{i\theta})$. Notice that $Z(e^{i\theta})$ admits an \emph{analytic extension} as a rational function
\begin{equation}
	Z(z) = \frac{1}{1 - e^{-\alpha T} z^{-1}} - \frac{1}{2},\quad z\in\Cbb,
\end{equation}
so that ${\Phi}(z)$ in \eqref{Phi_posi_real} is rational as well. Then after some straightforward calculations, we obtain a transfer function
\begin{equation}\label{eq:Expo-W}
W(z)=\frac{(1-e^{-2\alpha T})^{\frac{1}{2}}}{1-e^{-\alpha T} z^{-1}}
\end{equation}
which is a \emph{stable and minimum-phase} spectral factor of $\Phi(z)$, namely
\begin{equation}
	{\Phi}(z) = W(z) \, W(z^{-1})
\end{equation}
where the identity is valid in a neighborhood of the unit circle.
We see that $W(z)$ corresponds to an AR model of order one that depends on the parameter $\alpha_\srm:=\alpha T$. 
In consequence, samples of the discrete random field $y_\srm(\xb)$ can be generated by filtering a white noise through $d$ cascaded AR$(1)$ models, one for each dimension. 
Obviously, the algorithm achieves a very low computational cost because each AR component has only order one.
Compared with \cite{li2019stepwise}, our approach has a more concise frequency-domain interpretation and a neater time-domain implementation that avoids factorization of large matrices.
Moreover, the model can be reused for computing multiple realizations of the random filed because the filter $W(z)$ remains unchanged with a given covariance function when the sampling distance is fixed. Although the Cholesky factor in the stepwise CMD method \cite{li2019stepwise} can also be reused, our method requires much less storage since only the filter coefficients need to be stored.
In addition, our model can be used to compute samples of an \emph{arbitrary} size. On the contrary, the CMD has to be redone from the start if the sample size changes.
A comparison of the computational complexity between different algorithms will be given at the end of this section.


\subsection{Solution for the Gaussian covariance function: an ARMA model}

Unlike the case with an exponential covariance function, the problem with a general covariance function may not have an exact analytic solution. 
In this subsection, we discuss the case with an Gaussian covariance function in \eqref{eq:Gaussian function}. The corresponding sampled version is
\begin{equation}\label{sampled_cov_Gauss}
\rho_{\srm}(k) := \rho(k T) = \sigma^2 e^{-\alpha T^2 |k|^{2}},\quad k\in \Zbb.
\end{equation}
It is well known that the (continuous-time) Fourier transform of a Gaussian function is another Gaussian function which is certainly nonrational. We speculate that the same happens for the discrete samples of a Gaussian function, i.e., the corresponding spectral density is \emph{not} a rational function. Therefore, an ARMA representation for such a covariance function must by nature be approximate.


The Gaussian function \eqref{eq:Gaussian function} decays fast as $|x|$ increases. So a natural idea is to construct a rational spectral density 
\begin{equation}\label{eq:rational spectral}
\Phi(e^{i\theta})=\frac{P(e^{i\theta})}{Q(e^{i\theta})}
\end{equation}
that matches a \emph{finite} number of low-order (dominant) covariances, 
where $P$ and $Q$ are positive symmetric trigonometric polynomials. 
There are a number of solution techniques for this \emph{rational covariance extension} problem, see e.g., \cite{byrnes2001finite,Georgiou-L-03,RFP-09,FMP-12,Z-14,Z-14rat,ZFKZ2020M2-SIAM}. In the paper, we adopt the following \emph{generalized maximum entropy} formulation \cite{byrnes2001finite,ringh2016multidimensional}: 
\begin{equation}\label{Maximum entropy}
	\begin{aligned}
		&\underset{\Phi>0}{\max}\ \int_{\Tbb} P(e^{i\theta}) \log \Phi(e^{i\theta}) \frac{\d \theta}{2\pi} \\
		&\mathrm{s.t.}\quad \sigma_k = \int_{\Tbb}e^{ik \theta}\Phi(e^{i\theta})\frac{\d \theta}{2\pi}\quad \forall k\in\Lambda,
	\end{aligned}
\end{equation}
where,
\begin{itemize}
	\item $P$ is a \emph{known} positive symmetric polynomial which is constructed from a \emph{given} factor $b(z):=\sum_{k=0}^{n} b_{k}z^{-k}$, i.e., $P(z)=b(z)b(z^{-1})$;
	\item the index set is defined as $\Lambda:=\{-m,\dots,-1,0,1,\dots, m\}$ such that $m$ is a user-specified positive integer,
	\item $\sigma_k$'s are the covariance data evaluated from the covariance function, namely $\sigma_k = \rho_{\srm}(k)$.
\end{itemize}
More precisely, we choose $m$ to be the smallest positive integer such that $\rho_{\srm}(k)$ is practically zero for all $k>m$. In other words, the approximation procedure takes into account the covariances with significant values and discards the rest.
The optimization problem \eqref{Maximum entropy} is convex and has a unique solution $\Phi=P/\hat{Q}$, where $\hat{Q}$ is the optimal solution of the dual problem
\begin{equation}\label{eq:dual-problem}
	\underset{Q>0}{\min}\ \innerprod{\sigmab}{\qb}-\int_{\Tbb} P(e^{i\theta})\log Q(e^{i\theta}) \frac{\d\theta}{2\pi},
\end{equation}
where $\qb:=\{q_k\}_{k \in\Lambda}$ are Lagrange multipliers, $\innerprod{\sigmab}{\qb}:=\sum_{k\in \Lambda}\sigma_k q_k$ denotes the inner product, and $Q(e^{i\theta}):=\sum_{k =-m}^m q_k e^{-i k\theta}$ is a symmetric trigonometric polynomial.
For technical details we refer readers to \cite{byrnes2001finite,ringh2016multidimensional}. 
The reason for choosing this formulation is that we want to obtain a rational $\Phi(e^{i\theta})$ of the form \eqref{eq:rational spectral} which is directly connected to the ARMA model via spectral factorization.
More precisely, the polynomial $a(z)=\sum_{k=0}^{m} a_{k}z^{-k}$ corresponding to the AR coefficients can be determined via the Bauer method \cite{sayed2001survey} for 
factoring $\hat{Q}(z) = \sum_{k=-m}^{m} \hat{q}_{k}z^{-k}$ where $\hat{q}_k$'s are the optimal Lagrange multipliers. 




In this way, once the rational spectral density $\Phi(z)$ and the filter 
\begin{equation}\label{W_z_ARMA}
W(z) = \frac{b(z)}{a(z)} =\frac{\sum_{k=0}^{n} b_{k}z^{-k}}{\sum_{k=0}^{m} a_{k}z^{-k}}
\end{equation}
are constructed in each dimension, samples of the random field $y_\srm(\xb)$ can be generated in the same fashion as described in the previous subsection, now via the cascaded ARMA recursions. Each ARMA recursion has a fixed computational cost (though larger than that of the AR$(1)$ model) related to the model order $(m, n)$ which is chosen small. 

Next, we focus on the computational complexity of our approach to Problem~\ref{prob_realization}.
For practical applications, we are primarily interested in the $3$-d case. Assume that we are asked to generate samples of a random field with a size $\Nb=(N_1,N_2,N_3)\in\Zbb^3$, see \eqref{ind_set_RF}, and each ARMA model in the cascade has order $(m_j,n_j)$ for $j=1,2,3$. It is then common practice to generate samples of a slightly larger size $\Mb=(M_1, M_2, M_3)$ which can be taken as $(1+\beta)\Nb$, say with $\beta=0.1$, in order to reduce the ``transient'' effect of filtering caused by an artificial boundary condition.
For the exponential covariance function, the computational cost of the algorithm is proportional to the product $M_1M_2M_3$, so the complexity is $O(N_1N_2N_3)$ in which we have absorbed the constant $1+\beta$ into the capital $O$ notation.
For the Gaussian covariance function which corresponds to an ARMA model, the algorithm still mainly runs in $O(N_1N_2N_3)$ flops because the computational cost of solving a small-size convex optimization problem \eqref{eq:dual-problem} is far lower than that of implementing the ARMA recursion.

In Table~\ref{tab:table_Compu} we state the computational costs of different methods: traditional CMD by Cholesky decomposition, stepwise CMD, circulant embedding method, and our stochastic realization approach, where an instance with specific numbers is also shown for clarity.
\begin{table*}[t]
	\begin{center}
		\caption{\centering The computational complexity of different methods in the $3$-d case. In the particular example, we have $\Nb=(100,100,100)$, $\Cb=(512,512,512)$, and $\Mb=1.1\Nb$.}
		\label{tab:table_Compu}
		\begin{tabular}{lll}  
			\toprule   
			Methods& Major computational complexity&Example (flops) \\
			\midrule   
			CMD&$O(N_1^{3}N_2^{3}N_3^{3}) $ &$1.00\times10^{18}$\\
			Stepwise CMD &$O(N_1N_2N_3(N_1+N_2+N_3))$&$3.00\times10^8$\\
			Circulant Embedding&$O(C_1 C_2 C_3 \log_{2} (C_1 C_2 C_3))$ &$3.62\times 10^9$\\
			Stochastic Realization&$O(N_1N_2N_3)$&$1.00\times10^6$\\
			\bottomrule  
		\end{tabular}
	\end{center}
\end{table*}
It can be seen that our method has the lower computational complexity which is \emph{linear} in the number of samples.

\section{Application to multiscale simulations}\label{Sec:Multi-scale random fields}

In some applications, e.g., the analysis of certain geomaterial properties \cite{chen2012characterization,chen2016cpt}, we often face a challenging problem that \emph{refinements} of the generated samples of the random field are needed across several scales, where the spatial variability, i.e., the covariance function is expected to be maintained. 
Here by ``a fine-scale simulation'', we mean obtaining samples of the random field on a denser grid by \emph{interpolating} the already generated coarse-scale realization, instead of generating fine-scale samples from scratch. Our approach is particularly suitable for such a purpose as we shall describe shortly.


For simplicity, we assume that the sampling distance $T$ (which is the scale parameter) in \eqref{sampled_cov_exp} and \eqref{sampled_cov_Gauss} is halved in each fine-scale simulation, i.e., $T_1=\frac{1}{2}T$, $T_2=\frac{1}{2}T_1=\frac{1}{4}T$, where the subscript denotes the number of fine-scale simulations that we perform. We want to point out that the model parameters in \eqref{eq:Expo-W} and \eqref{W_z_ARMA} in general \emph{change} across different simulation scales since they depend on $T$, although the model order one is maintained in the case of an exponential covariance function. The change of the AR coefficients in \eqref{W_z_ARMA} is more implicit and comes from the fact that the covariance data $\sigma_k$'s are changed in a fine-scale simulation. In any case, once the fine-scale model \eqref{eq:Expo-W} or \eqref{W_z_ARMA} is constructed, the interpolation of the samples of the random field can be carried out by the same filtering technique as in the generation of the coarse-scale samples, \emph{when} suitable boundary conditions and the fine-scale noise input (which is not i.i.d.~any more) are provided. 

\subsection{The boundary conditions}

\begin{figure}[t!]
	\begin{center}
		\includegraphics[width=10cm]{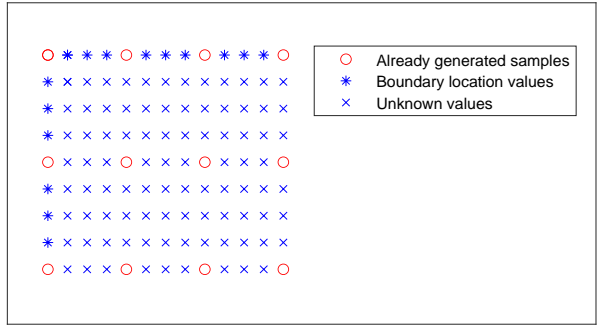}    
		\caption{A schematic figure for the boundary conditions in a multiscale simulation, where the number of values in boundary locations represents only a small fraction of the total number of unknown variables.}
		\label{Fig:2-d-multi-schematic-figure}
	\end{center}
\end{figure}

The determination of such boundary conditions is rather standard.
Suppose that the coarse-scale samples of the random field are collected into a column vector $\yb_1$ and the \emph{boundary} random variables that are needed to initiate the fine-scale filtering are collected in $\yb_2$. 
Here we want to remark on the special $1$-d case with the exponential covariance function, where the coarse-scale samples are themselves boundary conditions for the fine-scale simulation due to the AR$(1)$ model structure. Hence, all one needs to do is to generate the white noise samples on the new grid points and to implement the filtering. In general however, some boundary values of the fine-scale samples in $\yb_2$ are needed in order to start the multidimensional filtering, as shown in Fig.~\ref{Fig:2-d-multi-schematic-figure} for the $2$-d case where the two cascaded filters are both AR$(1)$.

Let $\yb$ be the joint vector such that
\begin{equation}\label{condi_normal}
\yb=\bmat \yb_1\\\yb_2\emat\sim \Ncal(\bmat \mathbf{0}\\ \mathbf{0}\emat, \bmat\Sigmab_{11}&\Sigmab_{12}\\ \Sigmab_{21}&\Sigmab_{22}\emat),
\end{equation} 
where we have introduced explicitly the Gaussianness assumption for the random field, and the matrix on the right side is
the covariance matrix $\Sigmab$ of $\yb$ which is evaluated using the given covariance function\footnote{Recall that the locations of samples in $\yb$ are known.} $\rho(\taub)$ in \eqref{eq:decoupled form} and partitioned in accordance with $\yb_1$ and $\yb_2$.
Then the unknown random vector $\yb_2$ conditioned on $\yb_1=\ab$ still has a multivariate normal distribution $(\yb_2|\yb_1=\ab)\sim \Ncal(\bar{\mub},\bar{\Sigmab})$ where
\begin{equation}\label{eq:mean-vector}
	\bar{\mub}=\Sigmab_{21}\Sigmab_{11}^{-1}\ab,
\end{equation}
and
\begin{equation}\label{eq:correlation-matrix}
	\bar{\Sigmab}=\Sigmab_{22}-\Sigmab_{21}\Sigmab^{-1}_{11}\Sigmab_{12}.
\end{equation}
The matrix $\bar{\Sigmab}$ is known as the \emph{Schur complement} of $\Sigmab_{11}$ in $\Sigmab$. It is positive definite because so is $\Sigmab$. 
Thus, the unknown boundary values $\yb_{2}$ in a fine-scale simulation can be computed via:
\begin{equation}
\yb_{2}=R\eb+\bar{\mub},
\end{equation}
where the matrix $R$ constitutes a \emph{rank factorization} of $\bar{\Sigmab}$, namely $\bar{\Sigmab}=R R^\top$ which can be computed from the spectral decomposition plus truncation, and $\eb$ is an i.i.d.~standard normal vector.

\subsection{The noise input}

Obviously, samples on the fine-scale depend upon the white noise input which cannot be randomly generated any more because of the existence of the coarse-scale samples as interpolation conditions. Thus it is reasonable to \emph{determine} the white noise before implementing ARMA recursion, and we report a particular solution for the case with exponential covariance functions, namely,
	\begin{equation*}\label{Exponential-multi-scale}
		\rho_{\srm} (k_1,k_2) = \sigma^2 e^{-\alpha_1 T_1 |k_1|-\alpha_2 T_2 |k_2|}, \quad (k_1,k_2)\in\Zbb^2.
	\end{equation*}
Here we focus on $2$-d random fields which are of engineering interest in e.g., \citep{chen2016cpt,chen2012characterization}, also for the simplicity of presentation. In principle, a similar procedure should work for the higher dimensional cases but the calculations will necessarily be more complicated.


	\begin{figure}
		\centering
		\begin{tikzpicture}[node distance=2cm,auto,>=latex']
		\node[draw, coordinate]                       		(start)   {};
		\node[draw, right=of start, minimum size=2em]       (step 1)  {$W_1(z_1)$};
		\node[draw, right=of step 1, minimum size=2em]      (step 2)  {$W_2(z_2)$};
		\node[draw, coordinate, right=of step 2]			(end)	  {};
		\draw[->] (start)  -- node[above] {$w(s,t)$} (step 1);
		\draw[->] (step 1) -- node[above] {$y_1(s,t)$} (step 2);
		\draw[->] (step 2) -- node[above] {$y(s,t)$} (end);
		\end{tikzpicture}
		\caption{The two cascaded linear stochastic system with given white noise input.}
		\label{Fig:2-cascaded filter}
	\end{figure}
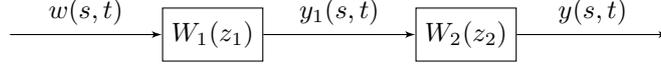

	Assume that we have computed the fine-scale samples in boundary locations, and we know the coarse-scale white noise and samples which have a size $(N_1,N_2)$.
	We aim to interpolate the fine-scale input $w(s,t)$ and then samples $y(s,t)$ for $0 \leq s \leq 2(N_1 - 1),\ 0 \leq t \leq 2(N_2 - 1)$. In particular, the indices $s$ and $t$ of the coarse-scale samples are \emph{even} numbers in the fine-scale realization.
	Referring to Fig.~\ref{Fig:2-cascaded filter}, the two cascaded AR(1) filters of the fine-scale random field are written as
	\begin{equation*}
	W_1(z_1)=\frac{b}{1-az_1^{-1}},\quad W_2(z_2)=\frac{d}{1-cz_2^{-1}}.
	\end{equation*}
	The input $w$, the intermediate output $y_1$, and the output $y$ are related via
	\begin{equation*}
	y_1(s,t)= W_1(z_1) w(s,t),\quad y(s,t)=W_2(z_2) y_1(s,t).
	\end{equation*}
	Apply twice the AR recursion, we have
	\begin{equation}\label{fine-scale w}
		w(s-1,t)=\frac{1}{ab} y_1(s,t) - \frac{a}{b} y_1(s-2,t) - \frac{1}{a} w(s,t).
	\end{equation}
    If we take $s = 2k$ and $t=2\ell$ with $k=1, \cdots, N_1-1$ and $\ell=1, \cdots, N_2-1$, $w(2k,2\ell)$ is known as the coarse-scale noise input. Then $w(2k-1,2\ell)$ can be computed directly via \eqref{fine-scale w} where
	\begin{equation*}
		y_1(2k,2\ell) = \frac{1}{d'} y(2k,2\ell) - \frac{c'}{d'} y(2k-2,2\ell),
	\end{equation*}
	where $c'$ and $d'$ correspond to the filter $W'_2(z_2)$ of coarse-scale random field, and the $y$ samples involved are from the coarse-scale realization.
	For odd $t=2\ell-1$, $w(2k,2\ell-1)$ is unknown and \eqref{fine-scale w} can be rewritten as a linear equation	
	\begin{equation}\label{linear-equation}
		A\wb = \bb
	\end{equation}
	with
	\begin{equation*}
		A=\bmat a&1&0&0&\cdots&0&0\\
		0&0&a&1&\cdots&0&0\\
		\vdots &\vdots &\vdots& \vdots& \ddots& \vdots&\vdots\\
		0&0&0&0&\cdots&a&1\emat,\,
		\wb=\bmat w(1,t)\\w(2,t)\\ \vdots\\ w(2N_1-2,t)\emat,\,
		\bb=\frac{1}{b} \bmat y_1(2,t)\\ y_1(4,t)\\ \vdots\\ y_1(2N_1-2,t) \emat - \frac{a^2}{b} \bmat y_1(0,t)\\ y_1(2,t)\\ \vdots\\ y_1(2N_1-4,t) \emat,
	\end{equation*}
	where the coefficient matrix $A$ has a size $(N_1 - 1) \times (2N_1-2)$, and $y_1 (2k,2\ell-1)$ is computed via
	\begin{equation*}
		y_1(2k,2\ell-1)= \frac{1}{cd} y(2k,2\ell) - \frac{c}{d} y(2k,2\ell-2) - \frac{1}{c} y_1(2k,2\ell),
	\end{equation*}
	where again the right-hand side requires only the coarse-scale samples.
    The equation \eqref{linear-equation} has infinitely many solutions since $A$ is of full row rank. In order to make the problem well-posed, we notice a \emph{dual} linear equation by swapping $W_1(z_1)$ and $W_2(z_2)$ in the filtering which obviously does not change the final output $y$. The intermediate output is, however, different and we write it as
    \begin{equation*}
    y_2(2k,2\ell)= \frac{1}{b'}y(2k,2\ell)- \frac{a'}{b'} y(2k,2\ell-2)
    \end{equation*}
    where $a'$ and $b'$ correspond to the filter $W'_1(z_1)$ of coarse-scale random field.
    We can now compute the components $w(2k,2\ell-1)$ in the vector $\wb$ as
	\begin{equation*}
		w(2k,2\ell-1)= \frac{1}{cd} y_2(2k,2\ell) -\frac{c}{d} y_2(2k,2\ell-2) -\frac{1}{c} w(2k,2\ell)
	\end{equation*}
    in the same fashion as computing $w(2k-1,2\ell)$ from \eqref{fine-scale w}. Substitute the result back into \eqref{linear-equation}, we get the rest components.

In this way, the fine-scale samples can be generated via the AR recursion easily as the values in the boundary locations and the white noise input have been computed.
We would like to point out that unlike traditional methods which generate \emph{all} the fine-scale samples (which could be a lot) using the conditional distribution \eqref{condi_normal}, see e.g., \citet[Sec.~2.4]{chen2012characterization}, 
we only need $\yb_2$ in boundary locations (whose number is much smaller, see Fig.~\ref{Fig:2-d-multi-schematic-figure}).
Therefore, our procedure involves multiplications and inversions of matrices of smaller sizes, which significantly improves computational efficiency. After the boundary values of the fine-scale random field are computed and the white noise input is obtained, the interpolation of the coarse-scale samples can be accomplished again at a linear cost.

\section{Numerical examples}\label{Sec:Numerical examples}

In this section, we perform two sets of numerical simulations of our stochastic realization approach: one for sampling $3$-d random fields and the other includes multiscale simulations in the $2$-d case.

\subsection{Sampling $3$-d Gaussian random fields}\label{Subsec:AR representation}

First, we apply the stochastic realization approach to Problem~\ref{prob_realization} with an exponential covariance function in three variables
\begin{equation}\label{eq:exponential-covariance-function}
	\rho_{\srm} (x,y,z)=\sigma^2 e^{-\alpha_1 T_1 |x|-\alpha_2 T_2 |y|-\alpha_3 T_3|z|},\quad (x,y,z)\in\Zbb^3.
\end{equation}
In the following example, the size of the samples to be generated is $\Nb = (100,100,100)$ for the random field $y_{\srm} (x,y,z)$, the variance is $\sigma^2 = 1$, the parameter vector is $(\alpha_1, \alpha_2, \alpha_3)=(1,1,1)$, and the vector of sampling distances along $x$-, $y$-, and $z$-directions is $(T_1, T_2, T_3) = (1/12, 1/10, 1/8)$.

\begin{figure}[t]
	\begin{center}
		\includegraphics[width=8.4cm]{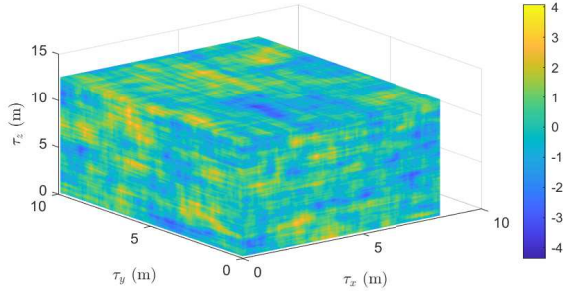}    
		\caption{A $3$-d random field realization with an exponential covariance function. The size of the realization is $8.33$m $\times 10.00$m $\times 12.50$m in space with a total number of samples equal to $100^3 = 10^6$.
		}
		\label{Fig:Three-Dimensional random field}
	\end{center}
\end{figure}
\begin{figure}[H]
	\centering
	\begin{minipage}[h]{0.48\textwidth}
		\centering
		\includegraphics[width=6cm]{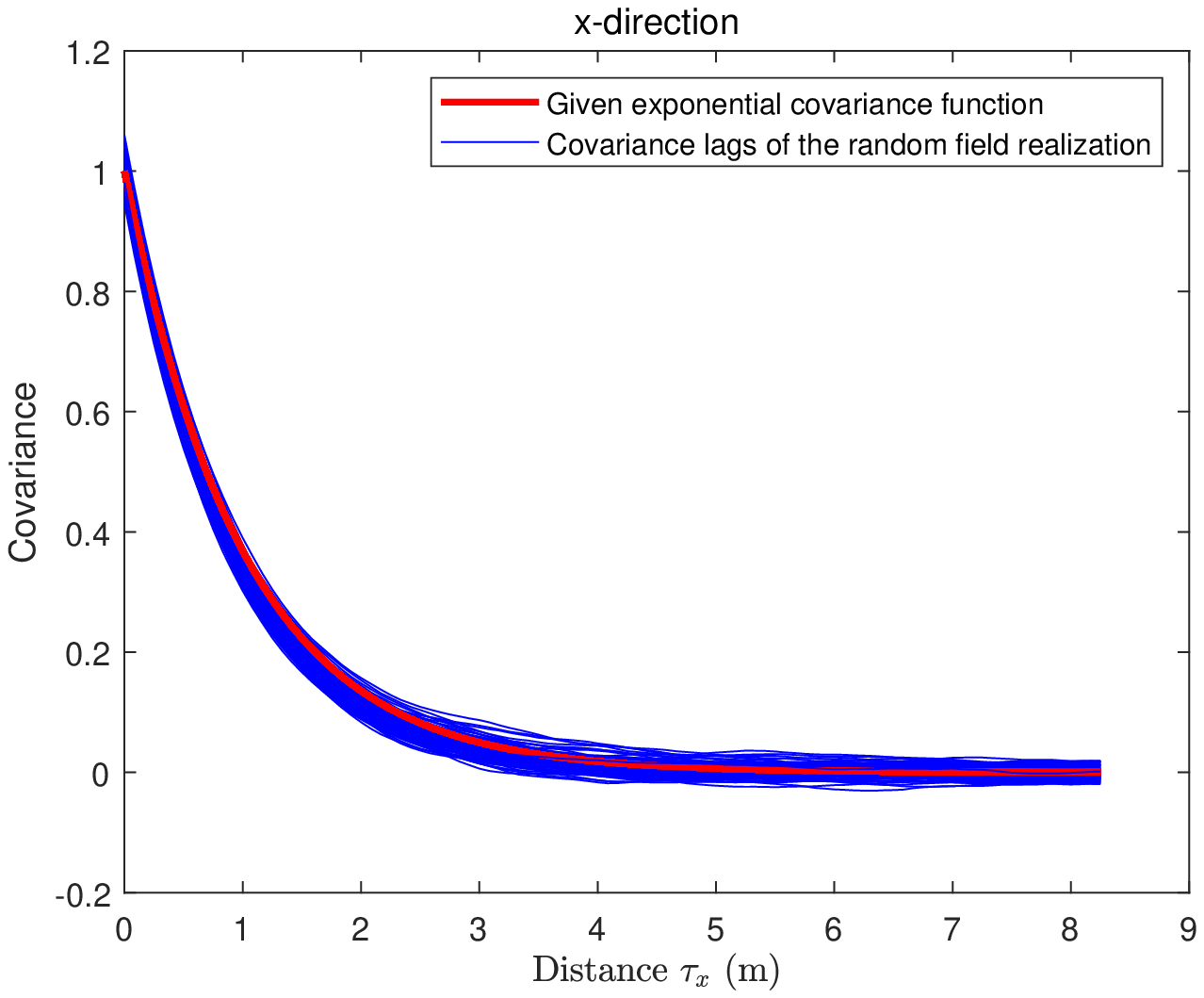}
	\end{minipage}
	\begin{minipage}[h]{0.48\textwidth}
		\centering
		\includegraphics[width=6cm]{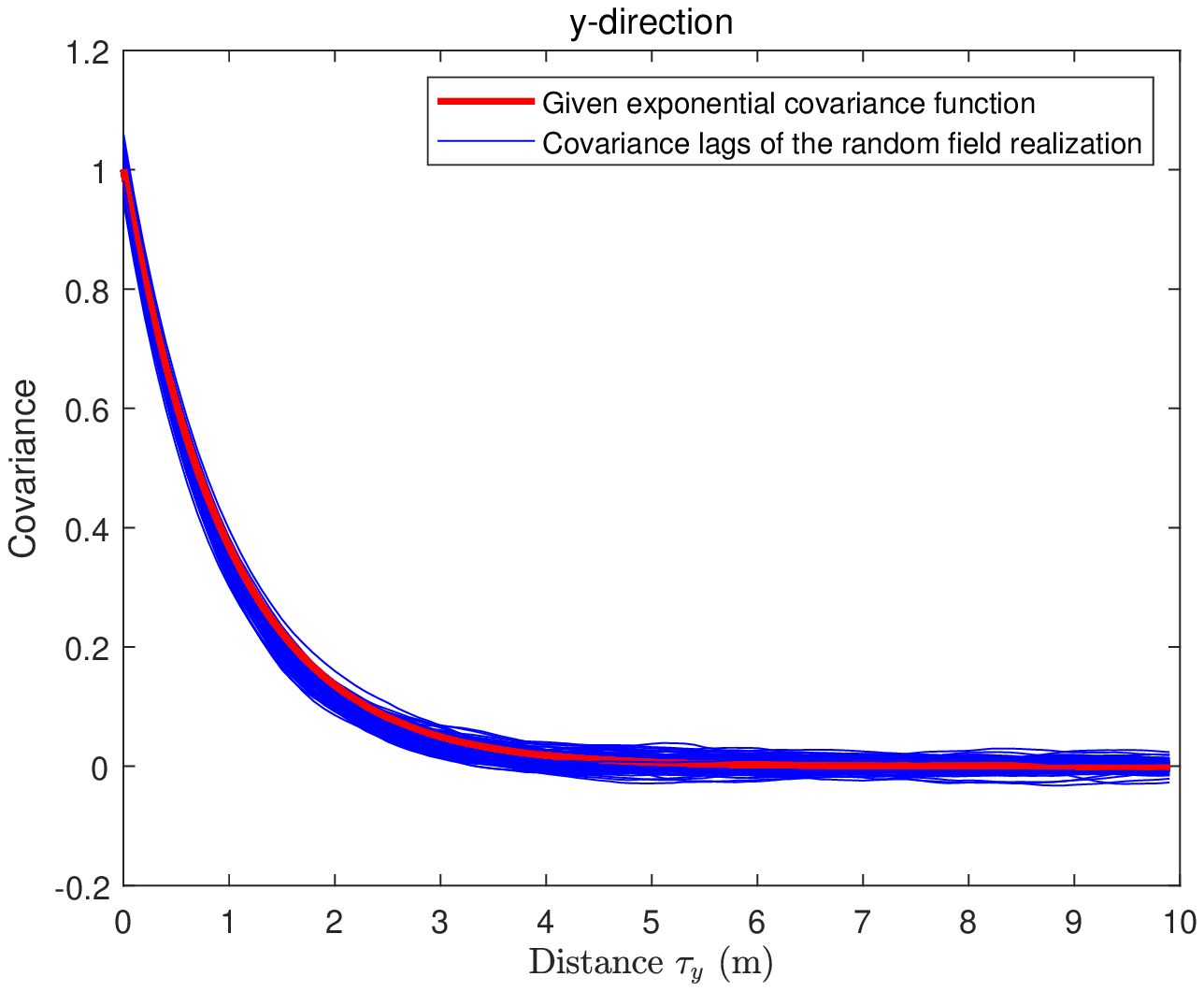}
	\end{minipage}
	\centering
	\begin{minipage}[h]{0.48\textwidth}
		\centering
		\includegraphics[width=6cm]{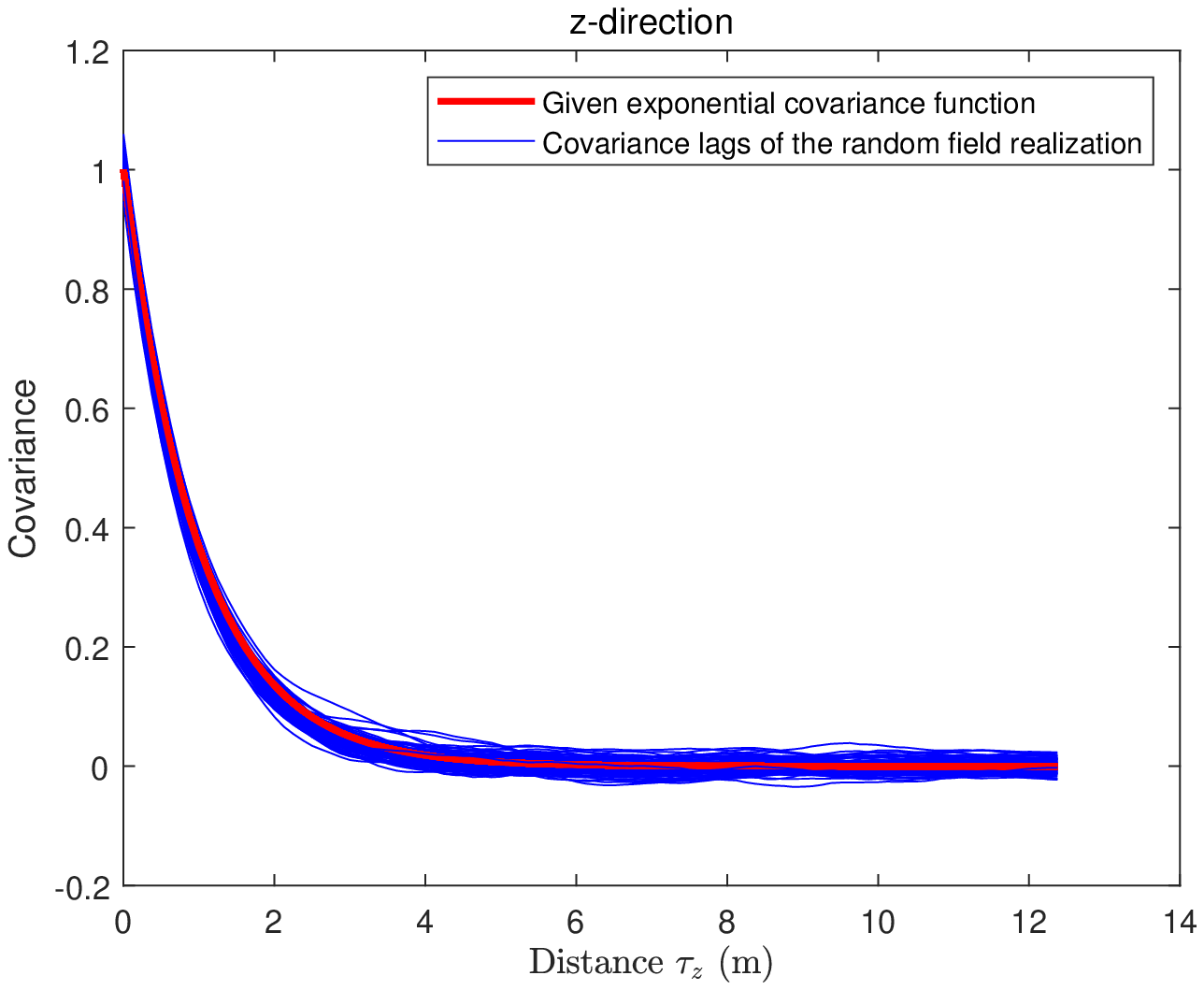}
	\end{minipage}
	\begin{minipage}[h]{0.48\textwidth}
		\centering
		\includegraphics[width=6cm]{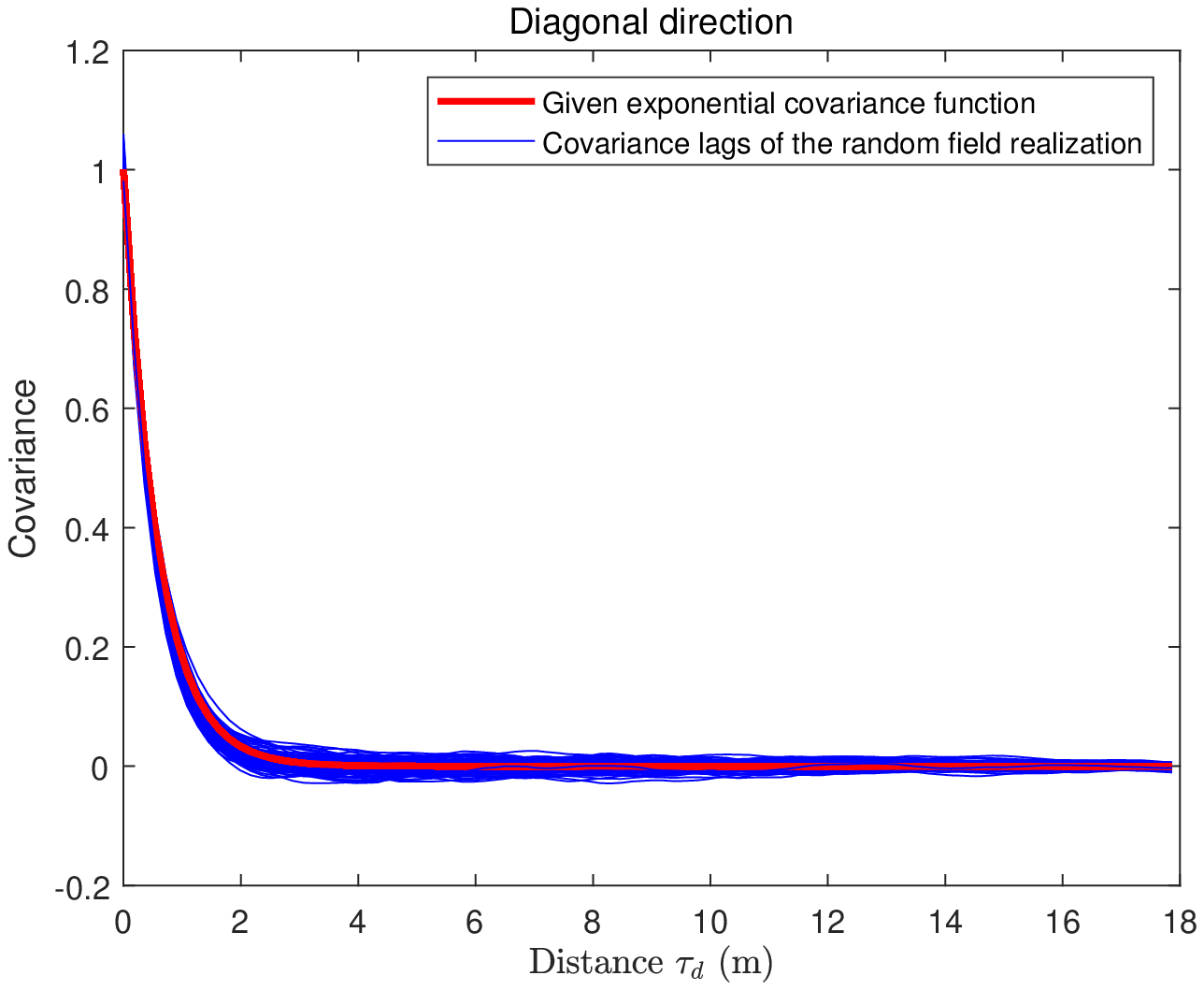}
	\end{minipage}
	\caption{The covariances versus distances along four directions of the $3$-d random field realization which contains 100 repeated trials. The red line denotes the given exponential covariance function and the blue lines are the corresponding sample covariance lags. By $x$-direction, we mean the section $[k_1,0,0]$ of the sample covariance array with $k_1=0,\dots,N_1-1$. The other directions are understood similarly.}
	\label{Fig:Four-direction covariance}
\end{figure}

It has been discussed in Subsection~\ref{subsec-expon} that the exponential covariance function corresponds to an exact rational spectral density, and the filter in each dimension can be directly computed via \eqref{eq:Expo-W}. 
Thus the required samples of the random field can be generated easily by implementing three cascaded AR recursions \eqref{d-dimensional ARMA model}. 
A realization of the random field is shown in Fig.~\ref{Fig:Three-Dimensional random field} using the Matlab command \verb*|slice|, where the spatial distances are defined as $\tau_x=T_1|x|$, $\tau_y = T_2|y|$, and $\tau_z = T_3|z|$ along three directions. 
Next, in order to verify the performance of our method, we also plot the \emph{sample covariances} of the realization $y_{\srm}(x,y,z)$ versus spatial distances along $x$-, $y$-, $z$-, and the diagonal directions in Fig.~\ref{Fig:Four-direction covariance} with $100$ repeated trials. 
The diagonal direction is along the line $x=y=z$ with $0\leq x \leq N_1-1$. 
\begin{figure}[t]
	\begin{center}
		\includegraphics[width=8.4cm]{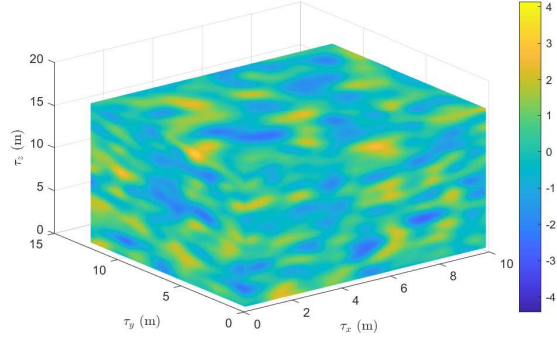}    
		\caption{A $3$-d random field realization with a Gaussian covariance function. The size of the realization is $2$m $\times$ $3.13$m $\times$ $5.56$m in space with a total number of samples equal to $50^3=1.25\times 10^5$.}
		\label{Fig:Three-Dimensional Gaussian random field}
	\end{center}
\end{figure}
\begin{figure}[H]
	\centering
	\begin{minipage}{0.48\textwidth}
		\centering
		\includegraphics[width=6cm]{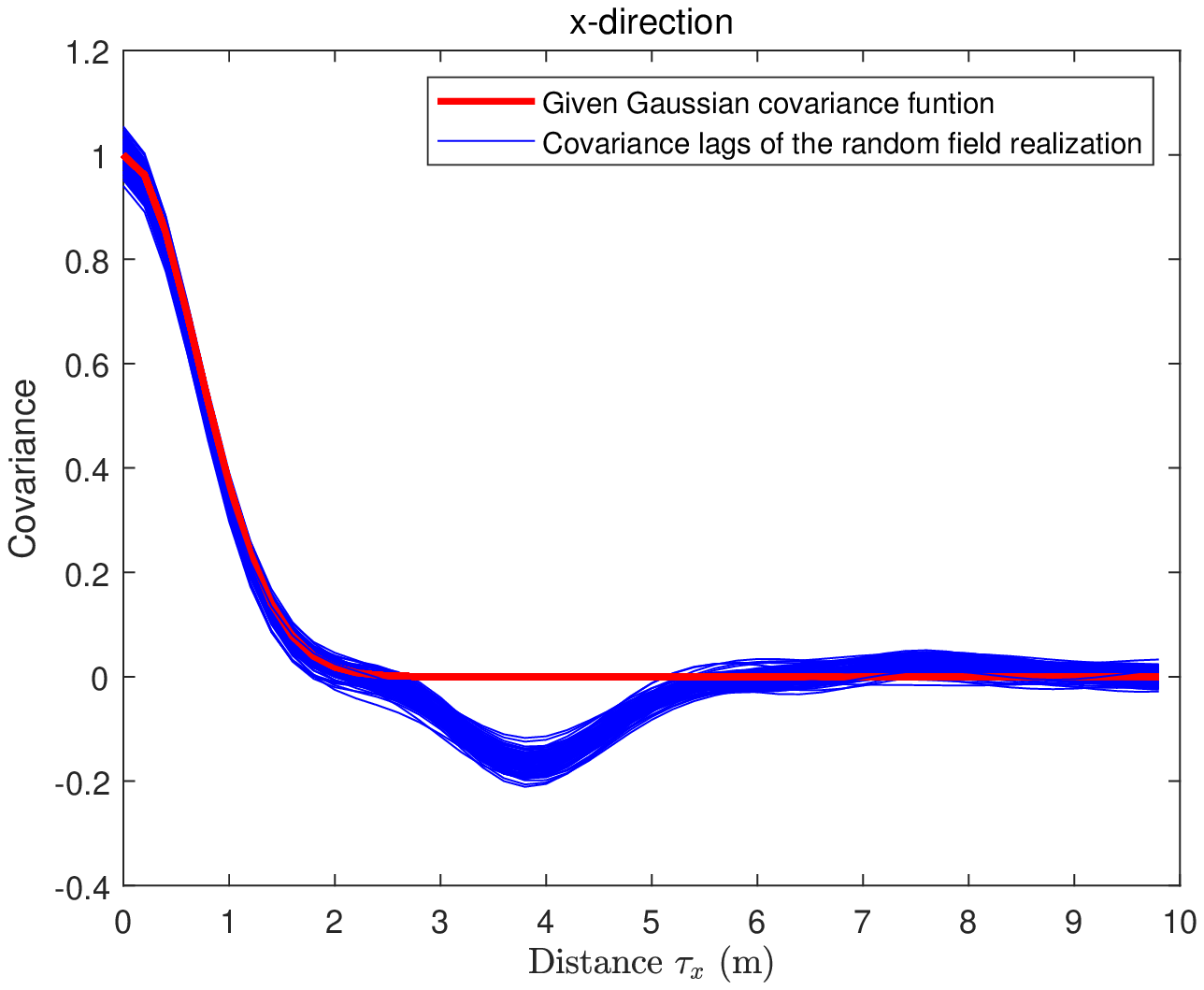}
	\end{minipage}
	\begin{minipage}{0.48\textwidth}
		\centering
		\includegraphics[width=6cm]{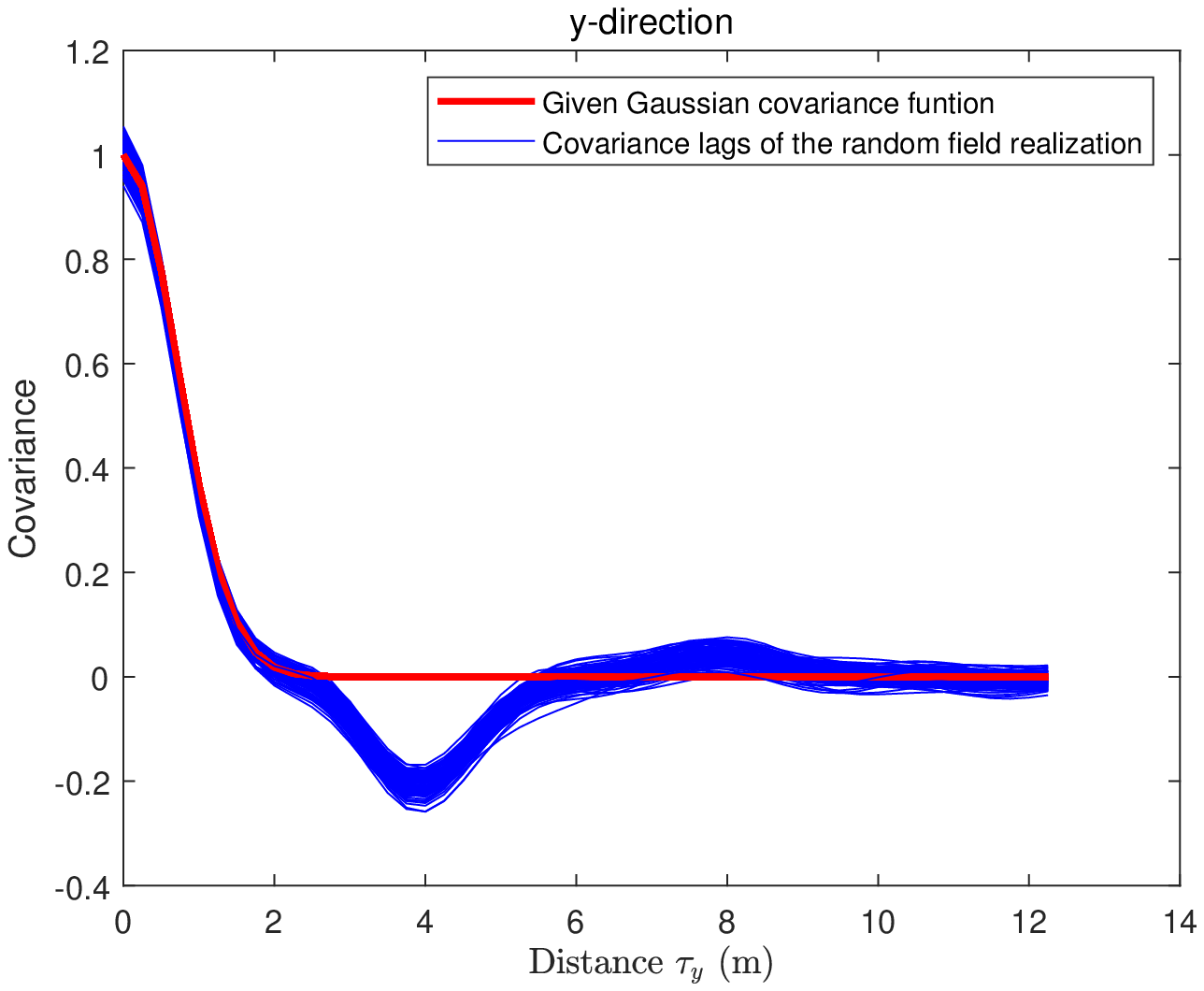}
	\end{minipage}
	\centering
	\begin{minipage}{0.48\textwidth}
		\centering
		\includegraphics[width=6cm]{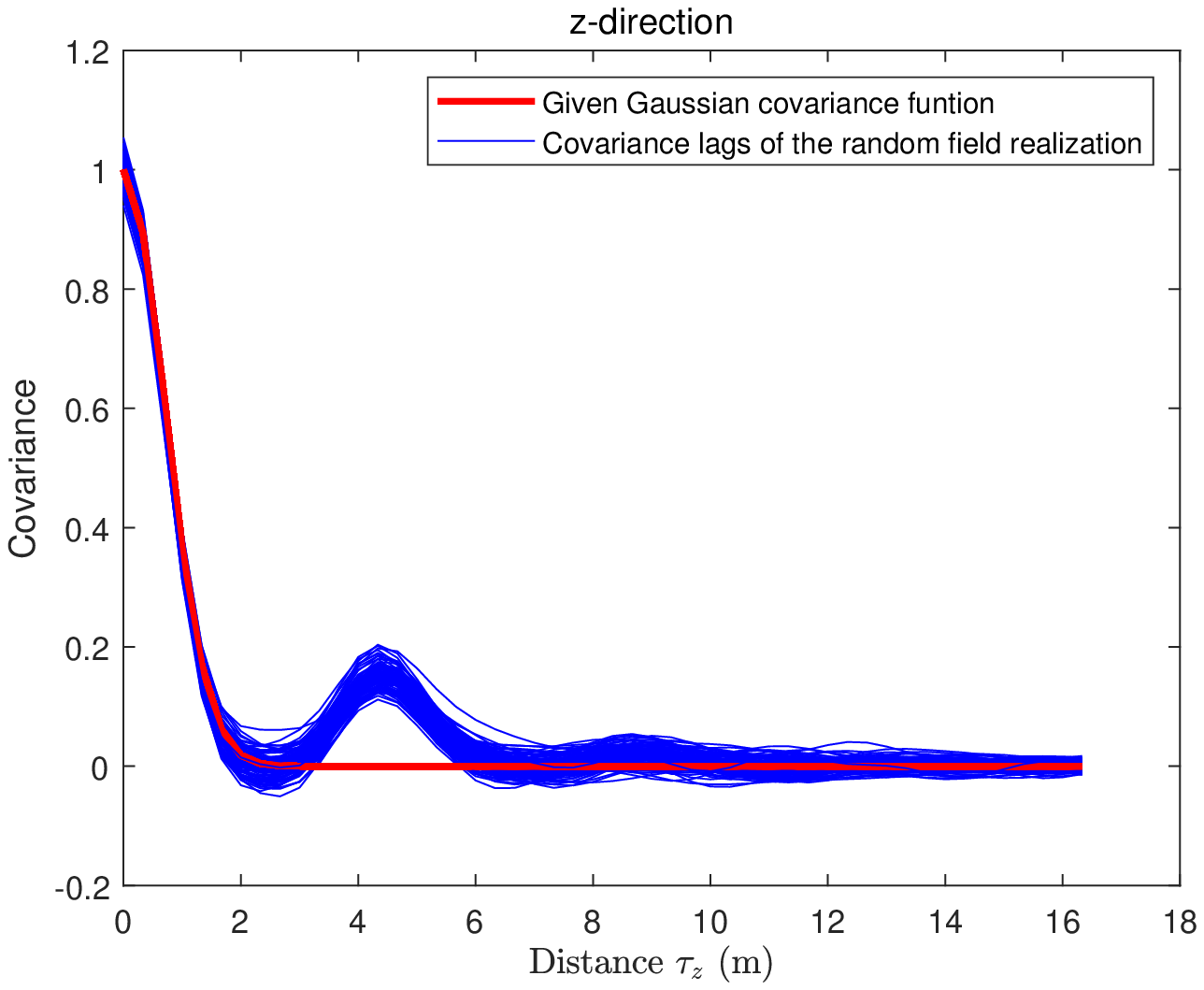}
	\end{minipage}
	\begin{minipage}{0.48\textwidth}
		\centering
		\includegraphics[width=6cm]{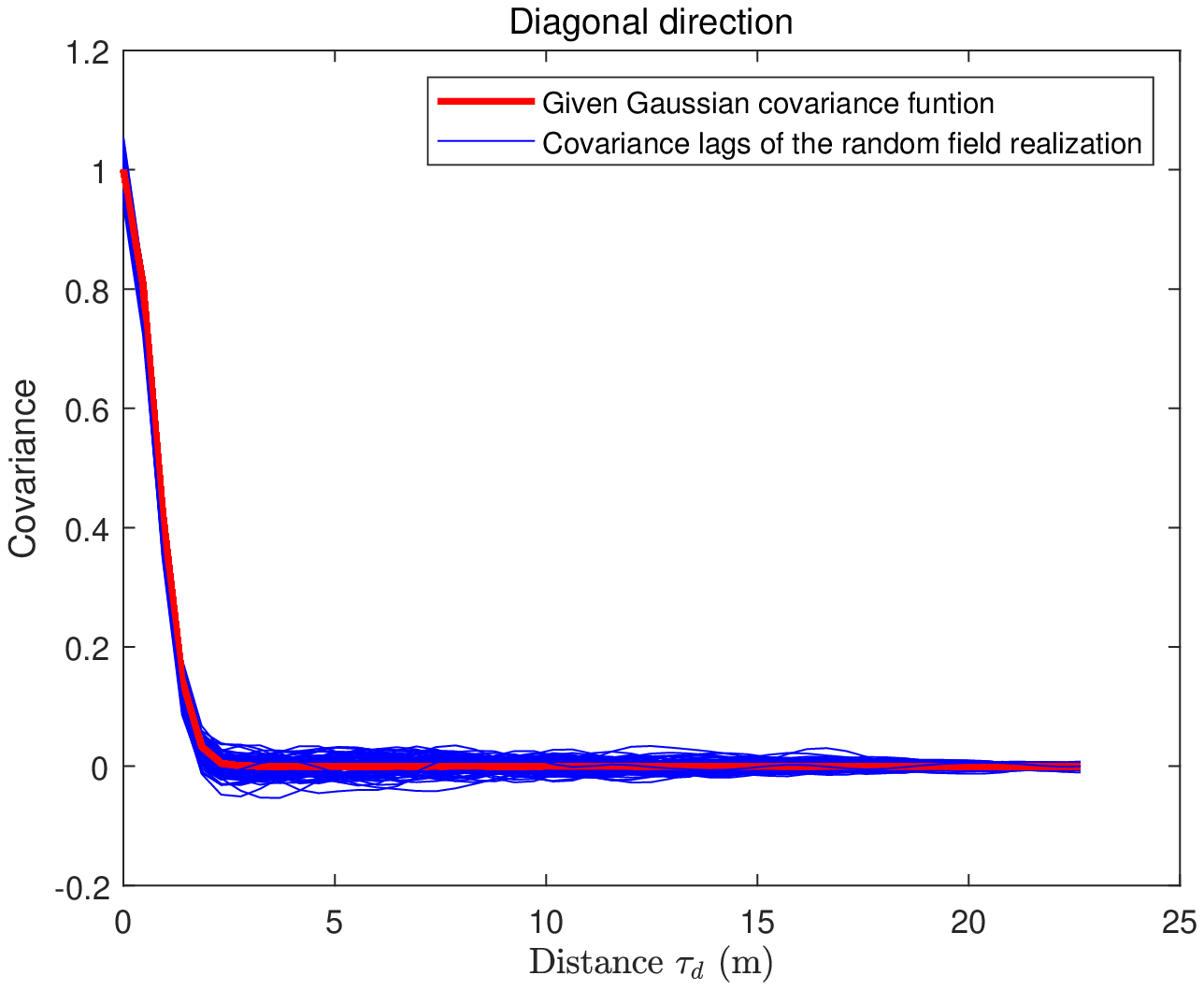}
	\end{minipage}
	\caption{The covariances versus distances along four directions of the $3$-d random field realization which contains 100 repeated trials. The red line denotes the given Gaussian covariance function and the blue lines are the corresponding sample covariance lags. By $x$-direction, we mean the section $[k_1,0,0]$ of the sample covariance array with $k_1=0,\dots,N_1-1$. The other directions are understood similarly.}
	\label{Fig:Four-direction covariance-Gaussian}
\end{figure}
In particular, the sample covariances of $y_{\srm} (x,y,z)$ are computed via the spatial average \cite[Section~5]{ZFKZ2-M2-Aut}:
\begin{equation}
	\hat\sigma_\kb:=\frac{1}{|\Nb|}\sum_{\xb} y_{\srm} (\xb+\kb){y_{\srm} (\xb)}	
\end{equation}
where $\xb=(x,y,z)$ and $|\Nb| = N_1 N_2 N_3$. 
Since we have explicitly enforced covariance matching in Problem~\ref{prob_realization}, it follows from the general covariance estimation theory \cite{priestley1981spectral} that the sample covariances of the output of the filter $W(\zb)$ must be close to the values of the given covariance function when the sample size is sufficiently large,
and this point is well illustrated in Fig~\ref{Fig:Four-direction covariance}. 
We remark that our result is visually much better than that reported in \cite[Sec.~5.1]{li2019stepwise}.


Furthermore, another simulation is performed for a Gaussian covariance function that has the form
\begin{equation}\label{eq:Gaussian-random-produce}
	\rho_{\srm} (x,y,z)=\sigma^2 e^{-\alpha_1 T_1^2 |x|^2- \alpha_2 T_2^2 |y|^2- \alpha_3 T_3^2 |z|^2}, \quad (x,y,z)\in\Zbb^3.
\end{equation}
The parameters are reported as follows: $\sigma^2=1$, $(\alpha_1, \alpha_2, \alpha_3) = (1,1,1)$, and $(T_1, T_2, T_3) = (1/5, 1/4, 1/3)$.
In this case, we set up an ARMA model to approximate the underlying spectral density. 
The numerator polynomial $b(\zb)$ in \eqref{eq:transfer_func} is specified by the user.
Here for simplicity, we take $b(z_1,z_2,z_3) = \prod_{j=1}^3 b_j(z_j)$ with a quite arbitrary $b_j(z_j)=1-0.2z_j^{-1}$ of order one and identical for $j=1,2,3$.
The orders of the denominator polynomials $a_j(z_j)$ are chosen to be $(m_1,m_2,m_3)=(8,7,7)$ which is a threshold for ``dominant'' covariances, i.e., large values of the covariance function. 
Then the approximate spectrum in each dimension is constructed via solving the optimization problem \eqref{Maximum entropy}, where Newton's method is implemented, and the polynomial $a_j(z_j)$ is obtained from spectral factorization.
The results are shown in Figs.~\ref{Fig:Three-Dimensional Gaussian random field} and \ref{Fig:Four-direction covariance-Gaussian} similar to the previous two figures.
Since the above Gaussian covariance function decays fast as the spatial distance increases, we set the sample size $\Nb=(50,50,50)$ in order to show more details in the figures. 
It can be seen that the sample covariances are very close to the given Gaussian covariance function when the lag is small. The mismatch for large lags can be explained by the fact that we  are using a relatively low-order ARMA spectrum to approximate the nonrational Gaussian function. 
In principle, we can also use a higher-order ARMA model for a better approximation, however, at the price of much more computational cost when generating the samples.


\subsection{Multiscale simulations in the $2$-d case}\label{Subsec:Multi-scale realization}

In this subsection, we perform simulations to produce fine-scale samples of the random field given the coarse-scale realization using our stochastic realization approach. Following the derivation in Section \ref{Sec:Multi-scale random fields}, we report an example with the following exponential covariance function
\begin{equation*}
		\rho_{\srm} (x,y)=\exp(-\frac{|x|}{5}-\frac{|y|}{4}), \quad (x,y)\in\Zbb^2.
\end{equation*}

\begin{figure}[t]
	\centering
	\begin{minipage}[h]{0.48\textwidth}
		\centering
		\includegraphics[width=7.8cm]{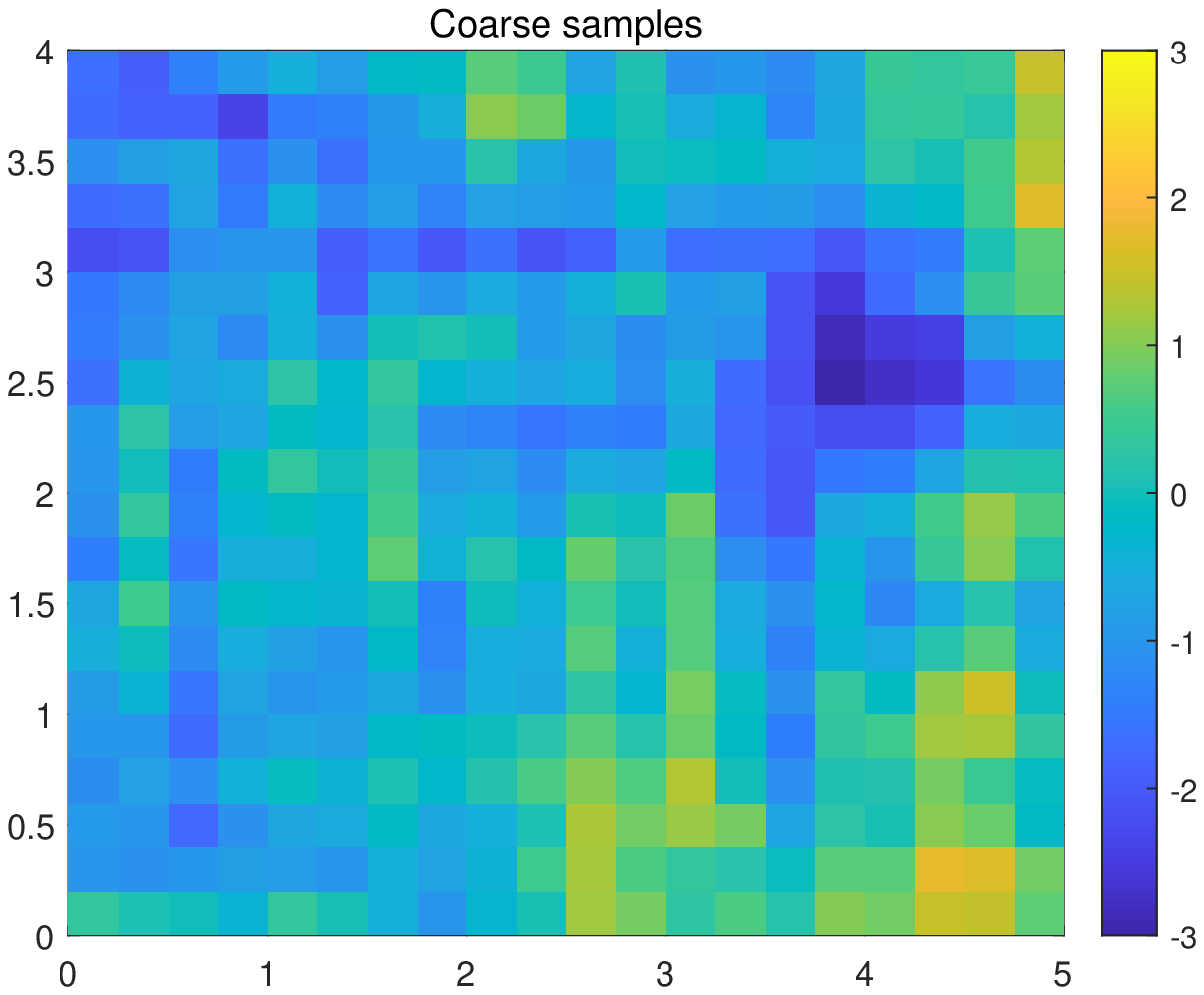}
		\centerline{(a) Coarse-scale samples ($20\times 20$)}
	\end{minipage}
	\begin{minipage}[h]{0.48\textwidth}
		\centering
		\includegraphics[width=7.8cm]{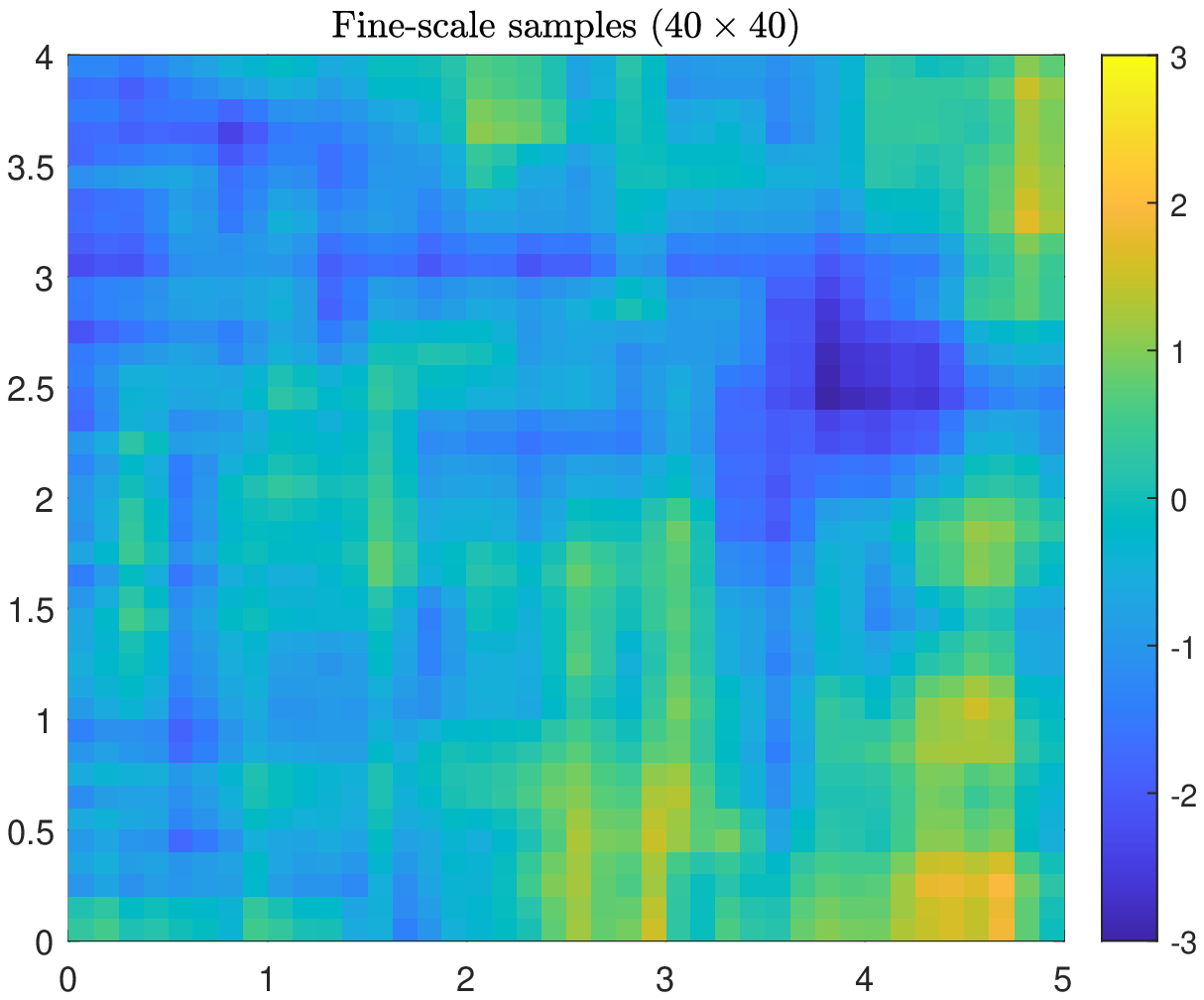}
		\centerline{(b) Fine-scale samples ($40\times40$)}
	\end{minipage}
	\centering
	\begin{minipage}[h]{0.48\textwidth}
		\centering
		\includegraphics[width=7.8cm]{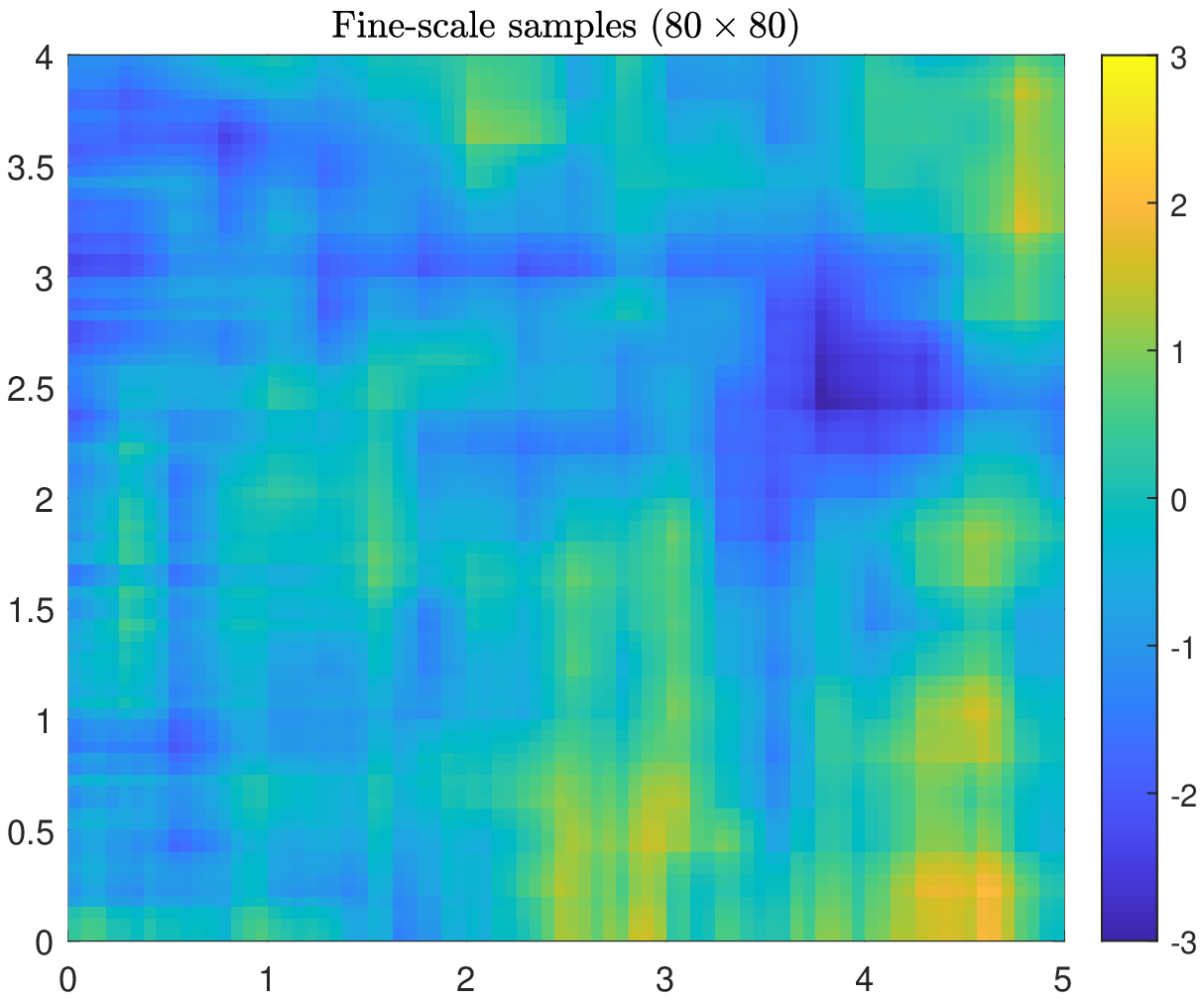}
		\centerline{(c) Fine-scale samples ($80 \times 80$)}
	\end{minipage}
	\begin{minipage}[h]{0.48\textwidth}
		\centering
		\includegraphics[width=7.8cm]{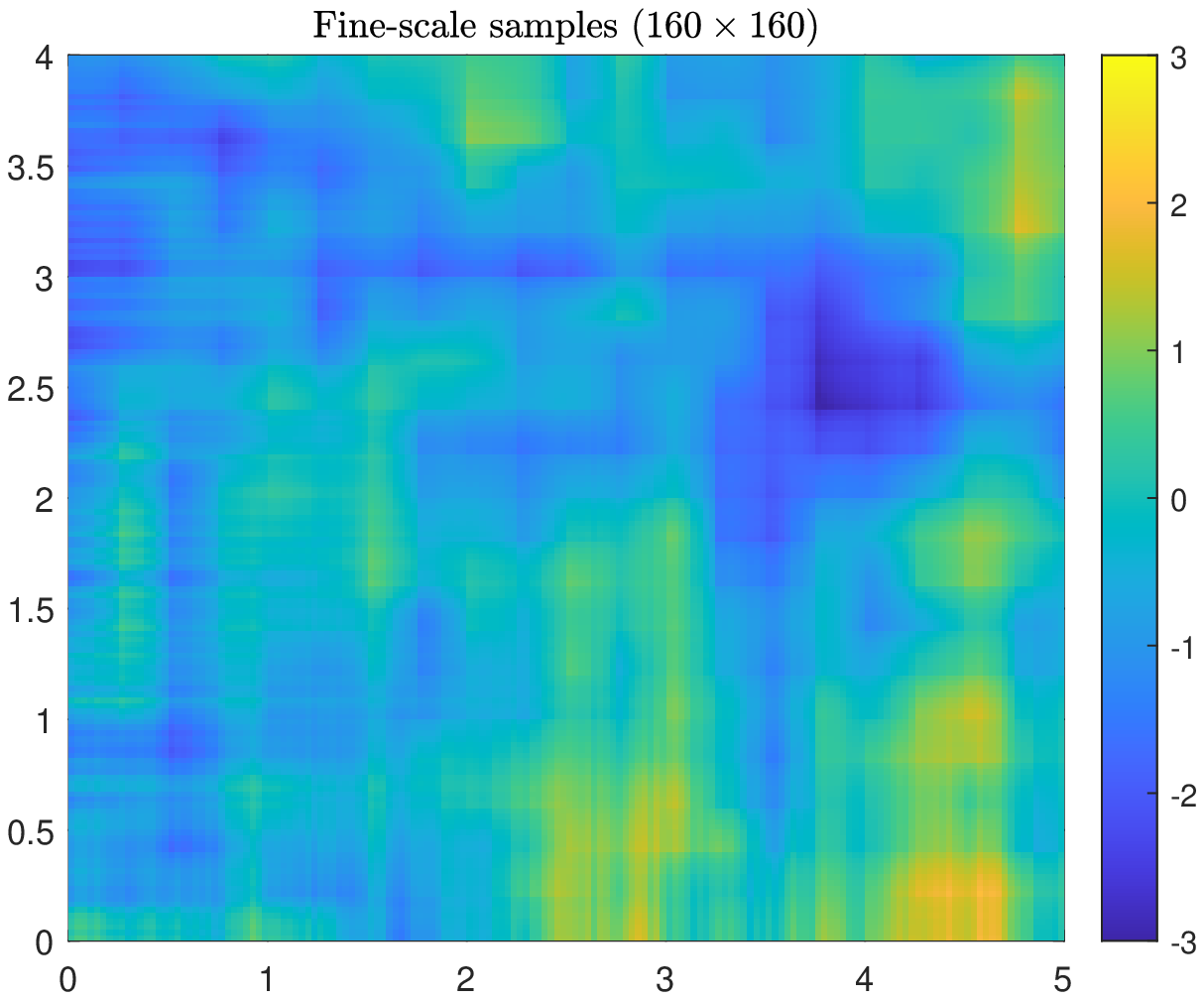}
		\centerline{(d) Fine-scale samples ($160\times 160$)}
	\end{minipage}
	\caption{Multiscale simulations with an exponential covariance function. Subfig.~(a) shows the coarse-scale samples (colored squares) which are already known, while Subfigs.~(b), (c), and (d) are the fine-scale realizations, where the numbers in parentheses give the number of sampling points in each direction.}
	\label{Fig:Figure:multi-scale}
\end{figure}

\begin{figure}[H]
	\centering
	\begin{minipage}[t]{0.48\textwidth}
		\centering
		\includegraphics[width=7.8cm]{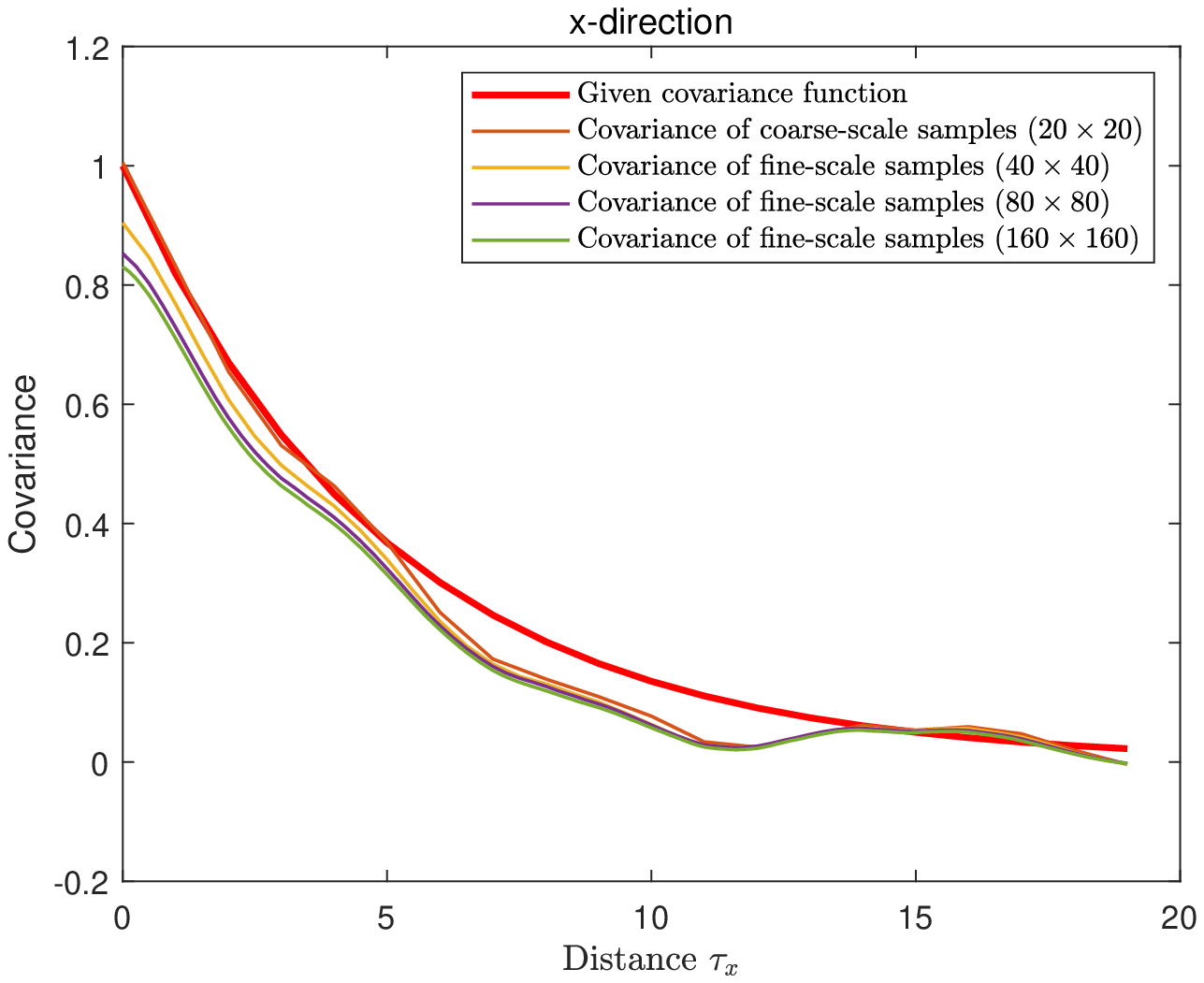}
	\end{minipage}
	\begin{minipage}[t]{0.48\textwidth}
		\centering
		\includegraphics[width=7.8cm]{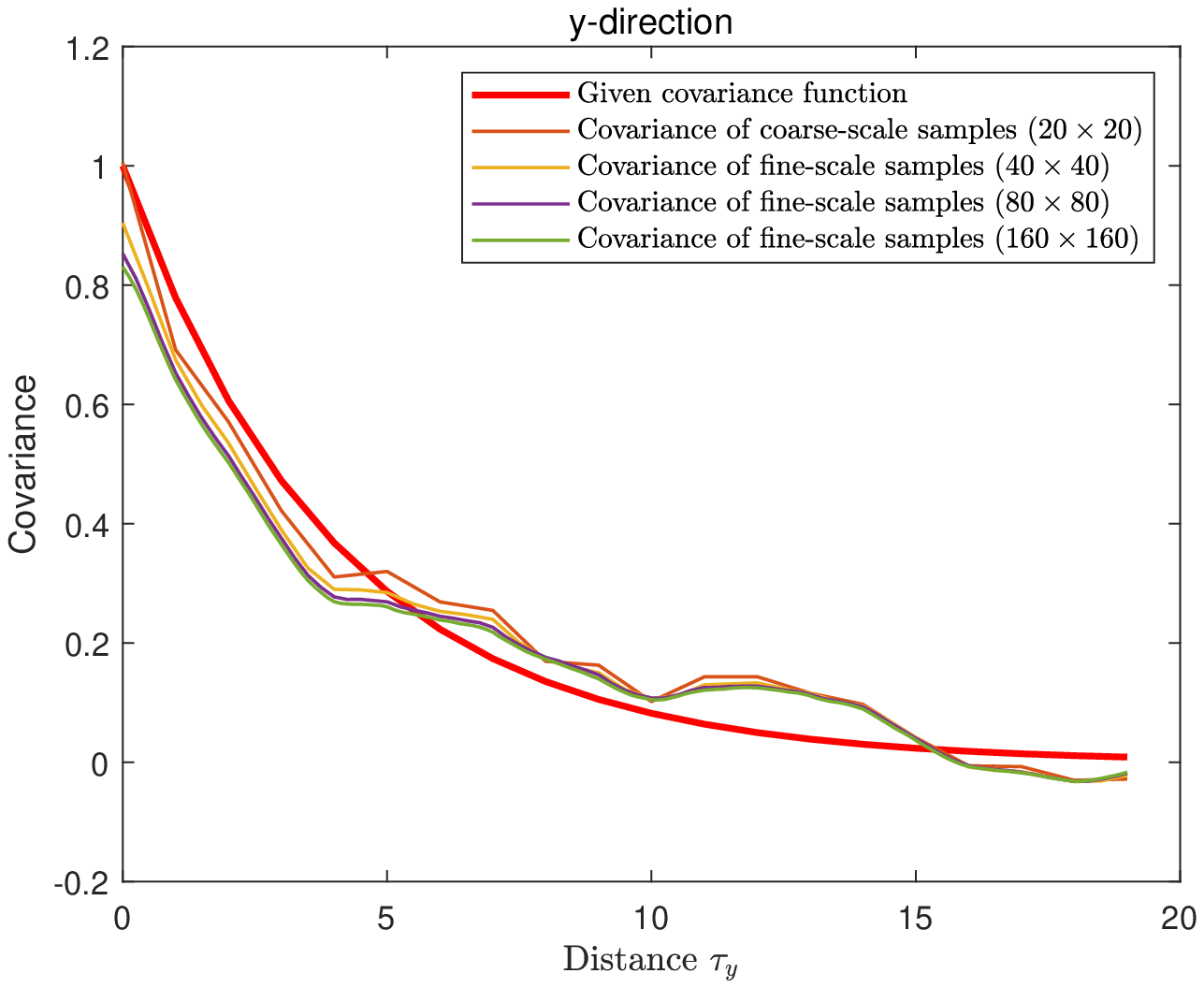}
	\end{minipage}
	\caption{Covariance lags of the random field versus distances. The red line denotes the given exponential covariance function, and the other four lines correspond to the sample covariances of under different scales. By $x$-direction, we mean the column of the sample covariance matrix indexed by  $[k_1,0]$ with $k_1=0,\dots,N_1-1$. The $y$-direction is understood similarly.}
	\label{Fig:Finer-scale random field Exponential}
\end{figure}

Suppose that the coarse-scale samples of the random field have a size vector $(N_1,N_2)=(20,20)$ which corresponds to $400$ points. We take the sampling distances to be $1/5$ and $1/4$ along $x$- and $y$-directions, respectively, i.e., $\Tb=(T_1,T_2)=(1/5,1/4)$, and the parameter vector $(\alpha_1, \alpha_2) = (1,1)$.
Then three fine-scale realizations of the random field are computed, where the sampling distances are reduced to $1/2$, $1/4$, and $1/8$ of the original values, respectively. 
The simulations are carried out sequentially. More precisely, after the samples with the parameter $\Tb$ are generated, they are then treated as the coarse-scale realization, and the interpolation procedure is executed to produce fine-scale samples with the parameter $\Tb'=\frac{1}{2}\Tb$.
Notice that we need to reconstruct the AR$(1)$ filter under each scale, due to the fact that the filter parameters in \eqref{eq:Expo-W}
depend upon the product $\alpha_j T_j$.
Then after the sample values in boundary locations are computed utilizing the conditional normal distribution and the white noise in unknown locations is obtained,
the rest fine-scale samples can be generated by implementing the AR recursion. These operations are repeated for each finer scale.
The result of such a multiscale simulation is shown in Fig.~\ref{Fig:Figure:multi-scale}. 
It is evident that more details of the random field can be seen from the fine-scale samples than the coarse-scale realization.
We also plot the sample covariances under each scale versus distances along $x$- and $y$-directions in Fig.~\ref{Fig:Finer-scale random field Exponential}. 
One can see that the sample covariances are again close to the given covariance function, which indicates that the spatial variability of the random field is well maintained across multiple scales in our simulations.

\section{Conclusions}\label{Sec:Conclusions}

This paper proposes an efficient stochastic realization approach for sampling large-scale multidimensional Gaussian stationary random fields.
The basic idea is to exploit the decoupling assumption on the covariance function, and to construct a rational model which approximates the spectrum of the underlying random field in terms of covariance matching.
Moreover, our sampling approach features easy implementation and 
{\setlength{\parindent}{0cm}
low computational complexity due to the simple structure of the approximate model.
The work in the paper can be concluded as follows:}
\begin{enumerate}
	\item[(1)] Solutions to the sampling problem with the exponential and Gaussian covariance functions are given, respectively. The former corresponds to a rational spectral density that leads to an AR$(1)$ filter, and the latter has a nonrational spectral density which can be approximated by an ARMA spectrum.	
	\item[(2)] The stochastic realization approach has been applied to multiscale simulations. Compared with traditional methods, only a few number of values in boundary locations are computed prior to the interpolation via the ARMA recursion, which achieves high efficiency.
	\item[(3)] Several numerical simulations are performed and they show that our method exhibits good performances not only in sampling large-size random fields, but also in refining generated samples across multiple scales. 
\end{enumerate}
Finally, it is expected that our approach can be extended to the multivariate case (i.e., vector processes), which will be a future study.

\section*{Acknowledgment}
This work was supported in part by the National Natural Science Foundation of China under the grant number 62103453 and the ``Hundred-Talent Program'' of Sun Yat-sen University.

\section*{Conflict of interest}
The authors declare no conflict of interest.

\bibliography{mybibfile}

\end{document}